\documentclass[smallextended]{svjour3}     % onecolumn (second format)
\smartqed  % flush right qed marks, e.g. at end of proof
\usepackage{amsfonts}
\usepackage{amsmath}
\usepackage{amssymb}
\usepackage{epsfig}
\usepackage{fullpage}
\usepackage{graphicx}
\usepackage{ifthen}
\usepackage{textcomp}

\newcommand{\area}{\operatorname{area}}

\newcommand{\ba}{{\bf a}}

\newcommand{\bbf}{{\bf f}}
\newcommand{\bby}{{\bf y}}
\newcommand{\bc}{{\bf c}}
\newcommand{\be}{\begin{equation}}

\newcommand{\beq}{\begin{equation}}
\newcommand{\beqn}{\begin{eqnarray*}}
\newcommand{\beqa}{\begin{eqnarray}}
\newcommand{\beqan}{\begin{eqnarray*}}
\newcommand{\blm}{\boldsymbol{\eta}}
\newcommand{\bma}{\boldsymbol{\alpha}}
\newcommand{\bme}{\boldsymbol{\eta}}
\newcommand{\bmo}{{\bf 0}}
\newcommand{\bo}{\boldsymbol}

\newcommand{\bp}{{\bf p}}
\newcommand{\br}{{\bf r}}
\newcommand{\bs}{{\bf s}}

\newcommand{\bx}{{\bf x}}
\newcommand{\bxi}{{\boldsymbol \xi}}
\newcommand{\bX}{{\bf X}}

\newcommand{\comment}[1]{}
\newcommand{\C}{\mathbb{C}}
\newcommand{\Caustic}{{\rm Caustic}}

\newcommand{\Crit}{{\rm Crit}}

\newcommand{\D}{\mathrm{d}}

\newcommand{\ee}{\end{equation}}
\newcommand{\eeq}{\end{equation}}
\newcommand{\eeqa}{\end{eqnarray}}
\newcommand{\eeqan}{\end{eqnarray*}}
\newcommand{\eeqn}{\end{eqnarray*}}

\newcommand{\f}{\mathbf{f}^\C_c}
\newcommand{\fo}{f^\C_1}
\newcommand{\ft}{f^\C_2}
\newcommand{\fix}{\mathrm{fix}}
\newcommand{\fkM}{\mathfrak{M}}

\newcommand{\frkS}{{\mathfrak S}}

\newcommand{\hol}{\mathrm{hol}}

\newcommand{\Hess}{\mathop{\rm Hess}\nolimits}

\newcommand{\I}{\mathrm{i}}

\newcommand{\Jac}{\mathop{\rm Jac}\nolimits}

\newcommand{\Mag}{{\rm Mag}}
\newcommand{\one}{{\mathbf 1}}
\newcommand{\oli}{\overline}

\newcommand{\PR}{\mathbb{P}}

\newcommand{\rmax}{{\rm max}}
\newcommand{\rmin}{{\rm min}}
\newcommand{\rideal}{R/(\varphi(x))}
\newcommand{\rsad}{{\rm sad}}
\newcommand{\R}{\mathbb{R}}

\newcommand{\RR}{\mathbb{R}}
\newcommand{\rd}{{\rm d}}
\newcommand{\ru}{{\rm u}}
\newcommand{\rv}{{\rm v}}

\newcommand{\sfC}{{\sf C}}
\newcommand{\sff}{{\sf f}}

\newcommand{\sfk}{{\sf k}}

\newcommand{\tta}{{\tt a}}

\newcommand{\ttB}{{\tt B}}
\newcommand{\ttE}{{\tt E}}
\newcommand{\ttk}{{\tt k}}
\newcommand{\ttG}{{\tt G}}
\newcommand{\tth}{{\tt h}}

\newcommand{\ttt}{{\tt t}}
\newcommand{\ttw}{{\tt w}}
\newcommand{\ttz}{{\tt z}}

\journalname{GRG}
\begin{document}

\title{Mathematics of Gravitational Lensing: Multiple Imaging and Magnification
\thanks{AOP acknowledges the partial support of NSF grant DMS-0707003.}}
\titlerunning{Mathematics of Gravitational Lensing}

\author{A. O. Petters \and M. C. Werner}

\authorrunning{Petters and Werner}

\institute{A. O. Petters \at Department of Mathematics and 
Department of Physics, Duke University, Durham NC 27708, USA.
%Tel.: +1-919-660-2812\\
%Fax: +1-919-660-2821\\
\\
\email{petters@math.duke.edu}  
\and
M. C. Werner \at Department of Mathematics,
Duke University, Durham NC 27708, USA.
%Tel.: +1-919-660-2811\\
%Fax: +1-919-660-2821\\
\email{werner@math.duke.edu}}

\date{Received: date / Accepted: date}

\maketitle

\begin{abstract}
The mathematical theory of gravitational lensing has revealed many
generic and global properties.
Beginning with multiple imaging,
we review Morse-theoretic image counting formulas and lower bound results, and
complex-algebraic upper bounds  in the case of single
and multiple lens planes. We discuss recent advances in the
mathematics of stochastic lensing, discussing 
a general formula for the global expected number of minimum lensed images
as well as asymptotic formulas for the probability densities of
the microlensing random time delay functions,
random lensing maps, and random shear, and an asymptotic 
expression for the global expected number of micro-minima.
Multiple imaging
in optical geometry and a spacetime setting are treated.
We review global magnification relation
results for model-dependent scenarios and cover recent developments 
on universal local magnification relations for higher order caustics.
\end{abstract}

\section{Introduction}
\label{intro}

\subsection{Overview and Conventions}

Two important anniversaries related to gravitational lensing occurred in
2009: ninety years ago the first
observation of this effect was announced at a joint meeting of the Royal
Society
and the Royal Astronomical Society, as a successful test of Einstein's
new theory of gravity;
and thirty years ago Walsh, Carswell and Weyman reported the first
observation of an extragalactic
example of lensing. Especially since then, the subject has become a
thriving research field at the
interface of astronomy, theoretical physics and mathematics. Some
current research highlights on
astrophysical and cosmological applications of lensing have been
discussed earlier in this Special Issue,
as well as possible lensing tests of modified theories of gravity in the
spirit of the original corroboration of General Relativity.

Of course, it has also emerged that gravitational lensing theory
is a rich research area in its
own right within mathematical physics. This aspect can be approached
from three different directions: 
the widely used and astrophysically important
thin-lens, weak-deflection approximation;\footnote{The thin-lens, weak-deflection approximation
       is sometimes called the {\it impulse approximation}.}
optical geometry, which considers
the properties of spatial light rays, a simplification that also makes the method
applicable to astrophysically relevant models; and 
a full general relativistic spacetime method that studies
null geodesics.
These approaches have proved to be
mathematically quite rich, with
applications of singularity theory, differential topology, Lorentzian
geometry, algebraic geometry, and probability theory. 
In
this review article, we discuss recent work in this
direction on two aspects of the weak deflection limit, image counting
and magnification.

One of the most basic problems in gravitational lensing is the number of
images produced. Yet, already this apparently
simple question turns out to be difficult. Image counting results using
Morse theory and complex methods are
reviewed in Section~\ref{sec:IC} for the single lens plane case, and in
Section~\ref{sec:IC-kplanes} for multiple lens planes.
The expected number of images in stochastic lensing is discussed in
Section~\ref{sec:stoch-lensing}, in particular for the asymptotic
microlensing case of Section~\ref{sec:stoch-microlensing-asymp}. From the point of view of
optical geometry,
image multiplicity is also a global effect as outlined in Section~\ref{sec:mult-images-optgeom}. Finally, conditions for the occurrence of multiple
images in spacetime are summarized in Section~\ref{sec:mult-images-spacetime}. Going beyond
image number, finer information
on lensed images can be gained from the magnification. It turns out that
image magnifications obey global magnification relations
for certain lens models, and local magnification relations near
singularities up to and beyond codimension three, which are universal.
These results, which involve deep properties of singularities
and algebraic geometry, are
discussed in Sections~\ref{sec:glob-magrel},
\ref{sec:loc-magrel}, and \ref{sec:univ-magrel-0}, respectively.

\smallskip

\noindent
{\it Theorem Convention:}  Our criteria for deciding whether a
mathematical result is called a theorem
will be driven by its importance in advancing our understanding of the
physical and/or
mathematical aspects of lensing. 
\smallskip

\noindent
{\it Citation Conventions:} The first time a result from
a paper is mentioned, the name(s) of the author(s) are stated next to
the bibliographic reference.
Thereafter, citations of the same result have the bibliographic
reference without the names.
The authors of this article are cited as AOP for Petters and MCW for
Werner.

\noindent
{\it Miscellaneous Conventions:} All light sources are treated as
point-like.
The symbol $\equiv$ indicates that a definition is given.
Physical quantities are given in appropriate dimensionless forms.

\subsection{Notation for Some Basic Weak Deflection Lensing Concepts}

%{\centering
\begin{figure}
\centering
\includegraphics[width=0.5\textwidth, height=0.15\textheight]{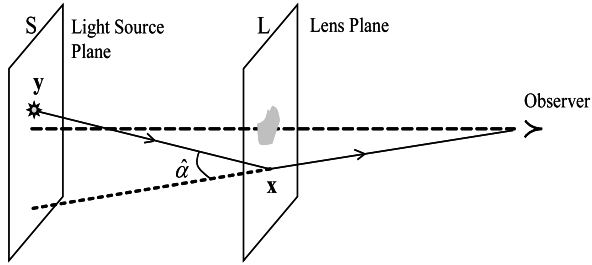}
\caption{A schematic of single-plane gravitational lensing. A pointlike light source
is at $\bby$ on the light source plane $S$.  A light ray from the source is deflected 
through an angle $\hat{\alpha}$ by 
the gravitational influence of the lens on the lens plane $L$.}
\label{fig:1plane-lensing}
\end{figure}
%}

Denote the dimensionless potential of a gravitational lens by $\psi$.
The (dimensionless) surface mass density $\kappa$ and 
magnitude $\Gamma$ of shear are:
$$
\kappa (\bx) = \frac{1}{2} \nabla^2 \psi(\bx),
\qquad
\Gamma^2 (\bx) = \frac{1}{4} \big(\psi_{\ru\ru} (\bx) - \psi_{\rv\rv} (\bx)\big)^2
+   \psi_{\ru\rv}^2(\bx),
$$ 
where $\bx = (\ru, \rv)$.  Note that $\bx$ is dimensionless, i.e., $\bx = \br/d_L$
with $\br$ the physical impact vector in the plane of the physical lens at
angular diameter distance $d_L$; see Figure~\ref{fig:1plane-lensing}.  Since
$d_L \gg |\br|$, we can treat $\bx$ as an angular vector.
A {\it singularity} of
    $\psi$ is a point $\ba \in \RR^2$ such that either $\psi(\bx) \rightarrow -\infty$
     or $\nabla^2 \psi(\bx) \rightarrow \infty$ as $\bx \rightarrow \ba$.
In particular, an {\it infinite singularity} of $\psi$ is
      a singularity $\ba$ for which $\psi (\bx) \rightarrow - \infty$ 
         as $\bx \rightarrow \ba$.
We shall call $\psi$ {\it nonsingular} if it has no singularities.
The potential $\psi$ is assumed to be smooth ($C^\infty$) everywhere on $\RR^2$, except 
on the set $A$ of singularities of $\psi$.

Let $T_\bby: L \rightarrow \RR$ be the (dimensionless) single-plane time delay function
induced by a lens potential $\psi$, where $L = \RR^2 - A$ with
$A$ the set of singularities
 of $\psi$:
$$
T_\bby (\bx) = \frac{|\bx - \bby|^2}{2} - \psi (\bx).
$$
Unless stated to the
contrary, we assume that $A$ is a finite set.
The point $\bby$ lies in the light source 
plane $S = \RR^2$; see Figure~\ref{fig:1plane-lensing}.
The light source plane $S$ is the set of all
      ``angular'' source positions $\bby = \bs/d_S$, where $\bs$ is a physical (linear) source position
       in the Euclidean plane at angular diameter distance $d_S$ and 
orthogonal to the line of sight.

By Fermat's principle \cite{SEF,PLW}, the light rays connecting $\bby$ to the
observer are given by the critical points of $T_\bby$, i.e.,
solutions $\bx \in L$ of
$$
\bmo = \nabla T_\bby (\bx) = - \bby + \bx - \nabla \psi (\bx),
$$
where the gradients are with respect to the rectangular coordinates $\bx = (\ru, \rv)$.
The above equation determines the single-plane lensing map corresponding to $T_\bby: L \rightarrow \RR$,
namely, the transformation $\bme: L \rightarrow S$ defined by:
$$
\bme (\bx) = \bx - \nabla \psi (\bx).
$$
The {\it lensed images} of a light source at $\bby$ are the elements of $\bme^{-1}(\bby)$
or, equivalently, the critical points of
$T_\bby$. Through this correspondence, we speak of minimum, saddle, and maximum
lensed images.
\smallskip

\noindent
{\bf Example (Microlensing):} We define microlensing generally as due to
a lens consisting of $g$ stars with masses $m_1, \dots, m_g$ 
 at respective positions $\bxi_1, \dots, \bxi_g$,
continuous matter with constant 
density $\kappa_c \ge 0 $, and an external shear $\gamma \ge 0 $.  
The lens potential of microlensing is then given by:
$$
 \psi_g(\bx)=\frac{\kappa_c}{2} |\bx|^2 -\frac{\gamma}{2}(\ru^2 - \rv^2)
       + \sum_{j=1}^g m_j  \log |\bx - \bxi_j |,
$$
where $\bx = (\ru, \rv)$.
The  induced time delay function $T_{g, \bby}$ at $\bby$ is 
$$
     T_{g,\bby}(\bx) =   \frac{1}{2} |\bx - \bby |^2-\frac{\kappa_c}{2} | \bx |^2 +\frac{\gamma}{2}(\ru^2-\rv^2)
       - \sum_{j=1}^g m_j  \log | \bx - \bxi_j |
$$
and the corresponding lensing map $\bme_g$ is 
$$
\bme_g(\bx) 
= \big( (1 - \kappa_c + \gamma) \ru,  (1 - \kappa_c - \gamma) \rv\big)
- \sum_{j=1}^g m_j \frac{\bx - \bxi_j}{|\bx - \bxi_j|^2}.
$$
We shall refer to the lensed images in microlensing as {\it micro-images} and
may even speak of {\it micro-minima} to designate minimum images in this context.
\smallskip

The ({\it absolute}) {\it magnification} of a lensed image is physically the ratio of the
flux of the image to the flux of the light source, which is given mathematically
by (e.g., \cite{PLW}, p. 85):
$$
\Mag(\bx; \bby) = \frac{1}{|\det[\Jac \bme](\bx)|}, \hspace{0.5in} \bme(\bx) = \bby.
$$
Note that
$
\det[\Jac \bme](\bx) = \det[\Hess T_\bby](\bx) =
\big(1- \kappa(\bx)\big)^2 - \Gamma^2 (\bx).  
$
The {\it signed magnification} of lensed image $\bx_i$ of $\bby$ is defined by:
$$
\mu_i = (-1)^{{\rm Ind}(\bx_i)} \Mag(\bx_i; \bby), \hspace{0.5in} \bme(\bx_i) = \bby.
$$
Here ${\rm Ind}(\bx_i)$ is the Morse index of $\bx_i$, i.e.,
the number of negative eigenvalues of $(\Hess T_\bby)(\bx_i)$,
which is $0$, $1$, and $2$ for
a minimum, saddle, and maximum, respectively.

The set $\Crit (\bme)$ of {\it critical points} of $\bme$ is the set of all
(formally) infinitely magnified lensed images for all light source positions in $S$,
which consists of all $\bx \in L$ where $\det[\Jac \bme](\bx) =0$.
When $Crit (\bme)$ consists of curves, we shall speak of the {\it critical curves}
of $\bme$.
 The
set of {\it caustics} of $\bme$ is 
$ \Caustic (\bme) = \bme (\Crit (\bme))\subset S$, which is the set of all light source positions
from which there is at least one infinitely magnified lensed image. 
For physically relevant settings, the set  $\Crit (\bme)$ is bounded, which
yields that  
 $\Crit (\bme)$ and
$\Caustic (\bme)$ are compact (\cite{PLW}, p. 293).

The single-plane time delay function 
$T_\bby$
is said to be  {\it subcritical at infinity} if for each non-caustic point $\bby$ and
$|\bx|$ sufficiently large, the
eigenvalues $\lambda_1 (\bx; \bby)$ and $\lambda_2 (\bx; \bby)$
of $\Hess T_\bby (\bx)$ are positive. This means that
\beq
\label{eq:pos-eigen}
\lambda_1 (\bx; \bby) = 1 - \kappa (\bx) + \Gamma (\bx) > 0,
\qquad
\lambda_1 (\bx; \bby) = 1 - \kappa (\bx) - \Gamma (\bx) > 0.
\eeq
 Condition (\ref{eq:pos-eigen}) implies:
$$
0 \le \kappa (\bx) < 1, \qquad 0 \le \Gamma (\bx) < 1.
$$
In addition, the time delay surface  (graph) of $T_\bby$ then has
positive Gauss curvature for $|\bx|$ sufficiently large.
When $T_\bby$ is subcritical at infinity and 
$T_\bby (\bx) \rightarrow \infty$ as $|\bx| \rightarrow \infty$, we call
$T_\bby$ {\it isolated}.

Given that the lensed images of a light source at $\bby$ are in 1-1 correspondence with
the  critical points of $T_\bby$, we shall characterize each lensed image as either
a minimum, saddle, or maximum.
\\

\noindent
{\bf Notation:}
    \begin{enumerate}
 \item $N =$ total number of images.
  \item $N_\rmin =$ number of minimum images.
  \item $N_\rsad =$ number of saddle images.
    \item $N_\rmax =$ number of maximum images.
   \item $N_+ = N_\rmin + N_\rmax = $ number of positive parity images.
   \end{enumerate}

\section{Multiple Imaging in Single-Plane Lensing}
\label{sec:IC}

\subsection{Image Counting Formulas and Lower Bounds: Single Plane Lensing}
\label{sec:IC-1}

Einstein determined in 1912 \cite{RSS97} that a lens consisting of a single
star will produce {\it two} images of  a background star
that is not on a caustic.
Two stars on the same lens
plane can produce either {\it three} or {\it five} images for
light sources off a caustic, a result found in 1986 by
Schneider and Weiss \cite{Sch-Weiss86}.
The method in \cite{Sch-Weiss86}, however, involved lengthy calculations that
directly solved the lens equation for the images, an approach that would be
impossible to carry over to any finite number of stars or more general
mass distributions. 

In 1991, AOP \cite{Ptt91} approached 
the image counting problem by employing Morse theory under boundary conditions
and generic properties of time delay functions 
to obtain a general theorem yielding counting formulas and lower bounds
for the number of images. 
The Morse theoretic approach is particularly powerful because it
gives specific counting information about the number of images of different types
and extends naturally to $k$-lens planes and a general spacetime setting. 
For simplicity, we state the results in \cite{Ptt91}
only for generic subcritical lensing (see \cite[Chap. 11]{PLW} for more):

\begin{theorem}{\rm \cite{Ptt91}}  
\label{thm:AOP91}
{\bf (Single-Plane Image Counting Formulas and Lower Bounds)}
Let  $T_\bby: L \rightarrow \RR$ be a single-plane time delay function
induced by a lens potential $\psi$ with $g\ge 0$ singularities, 
all of which are infinite singularities.
Suppose that $\bby$ is not on a caustic and $T_\bby$ is isolated. Then for a 
generic\footnote{Theorem~\ref{thm:AOP91} holds either for $T_\bby$ or
     almost all sufficiently small linear perturbations of $T_\bby$,
     i.e., the functions $T_\bby (\bx) + \bp\cdot \bx$ for every $\bp \in \RR^2$, 
      except in a set of measure zero, with $|\bp|$ sufficiently small \cite[p. 421]{PLW}.}
$T_\bby$ the number of lensed images obeys: 
\begin{enumerate}
 \item $N = 2 N_+ + g -1 = 2 N_\rsad - g + 1$, \qquad $N_+ = N_\rsad - g + 1$.
 \item $N \ge g + 1$, $N_\rmin \ge 1$, $N_\rsad  \ge N_\rmax + g$.
 \item If the corresponding lensing map is locally stable\footnote{Local stability is
 equivalent to the set of critical points 
        of $\bme$ consisting only of folds and cusps \cite[p. 294]{PLW}.}
with  $\Crit (\bme)$ bounded, then for sufficiently large $|\bby|$ the lower bounds on
    the number of images are attained:\\
       $N = g + 1$, $N_\rmin = 1$, $N_\rmax = 0$, and $N_\rsad = g$.
\end{enumerate}
\end{theorem}

%\noindent
As long as the hypotheses of Theorem~\ref{thm:AOP91} hold,
the image counting information
is independent of the choice of gravitational lens
model. The 
topological nature of the counting formulas and lower bounds give
them wide applicability.  Part 1 of Theorem~\ref{thm:AOP91} states that the number of
images has parity (even-ness or odd-ness) opposite to the number of singularities,
part 2 gives lower bounds---e.g., there are at least $g + 1$ images and at least
$g$ saddle images, and part 3 implies that the lower bounds in Part 2
are actually the smallest number of images that are achievable. 

\noindent
{\it Application to microlensing:} Theorem~\ref{thm:AOP91} applies to subcritical microlensing, i.e.,
lensing due to 
$g$ point masses with continuous matter $\kappa_c$ and shear $\gamma$ satisfying 
$1 - \kappa_c + \gamma > 0$ and
$1 - \kappa_c + \gamma > 0$ .  There is an even number of images if and only
if the number of stars is odd. Also, since there is no maximum images,
we have 
$$
N_\rmin = N_\rsad - g + 1.
$$
This is a useful formula for checking whether 
images are overlooked by in microlensing simulations.  Consult \cite[Chap. 11]{PLW} for a detailed discussion, where the
cases $1 - \kappa_c + \gamma > 0$ and
$1 - \kappa_c + \gamma < 0$ (strong shear lensing) and $1 - \kappa_c + \gamma < 0$ and
$1 - \kappa_c + \gamma < 0$ (supercritical lensing) are also treated. 
\smallskip

In the situation of a nonsingular lens,
the following corollary of Theorem~\ref{thm:AOP91}(1) recovers (by setting $g = 0$)
the single-plane Odd Number Image Theorem
found in 1981 by Burke \cite{Bur81}, who proved the result
using a different approach, namely, the Poincar\'e-Hopf index theorem.

\begin{corollary}{\rm \cite{Bur81}} \label{cor:AOP91}
{\bf (Single-Plane Odd Number Image Theorem)}
For a non-caustic point $\bby$, let $T_\bby$ be a single-plane isolated time delay function 
induced by a nonsingular gravitational lens potential. Then the 
total number of images is odd:
$$
N = 2 N_+ -1 = 2 N_\rsad + 1.
$$
\end{corollary}

  Nonsingular lensing is typically due to
modeling galaxies as smooth on a macro scale.  Though the predicted number
of images is odd, typically an even number of images is observed.
The reason is that maximum images are angularly located where the
surface mass density of the lens is supercritical
(core of galaxies), which causes them to become 
significantly demagnified (e.g., \cite{NA86a}, \cite[p. 470]{PLW}).

We saw from Theorem~\ref{thm:AOP91}(2)  that if an isolated lens has at least one singularity
($g\ge 1$), which can be a point mass, singular isothermal sphere, etc.,
then the lens can produce multiple images. Namely, there is a point
in the light source plane from which a light source has more than one lensed
image, $N > 1$.  How about multiple imaging due to an isolated nonsingular lens?
The following necessary and sufficient condition for multiple imaging
by such lenses
was established in 1986 by Subramanian and Cowling \cite{SubC86}:

\begin{theorem}{\rm \cite{SubC86}}
\label{thm:SubC86}
{\bf (Criterion for Multiple Images: Single-Plane Nonsingular Case)}
For a non-caustic point $\bby$, let $T_\bby$ be a single-plane isolated time delay function
induced by a nonsingular gravitational lens potential. Then:
\begin{enumerate}
 \item  $N \ge 3$ if and only if there is a point $\bx_0$ in the lens
plane $L$ such that 
$
\det[\Hess T_\bby](\bx_0) < 0.
%= (1 - \kappa (\bx_0))^2 - \Gamma^2 (\bx_0) < 0.
$
 \item If  there is a point $\bx_0\in L$ where the 
     surface mass density is supercritical, $\kappa (\bx_0) > 1$,
     then a light source at $\bby_0 = \bme (\bx_0)$ will have
      multiple images, $N \ge 3$. 
     \end{enumerate}
\end{theorem}

\noindent
We prove Theorem~\ref{thm:SubC86} to illustrate how
some of the previous counting results and ideas are used theoretically.
\begin{proof}
(1) If $N > 1$, then Corollary~\ref{cor:AOP91}(1) yields $N = 2 N_\rsad + 1 \ge 3$,
so $N_\rsad \ge 1$. Because there is at least one saddle image, say, $\bx_{\rm sad}$,
the point $\bx_{\rm sad}\in L$ satisfies $\det[\Hess T_\bby](\bx_{\rm sad}) < 0$.
 Conversely, if there is a point $\bx_0 \in L$ such that 
$\det[\Hess T_\bby](\bx_0) <0$, then a light source at $\bby_0 = \bme (\bx_0)$
has at least one saddle image.  Since $N_\rsad \ge 1$, $N_\rmin \ge 1$,
and the number of images is odd, we have $N \ge 3$.
\newline
(2) If  $\kappa (\bx_0) > 1$, then image $\bx_0$ cannot be a minimum since
minima are located where the surface mass density is subcritical (e.g.,
\cite[p. 423]{PLW}).  Then $\bx_0$ is either a saddle or maximum.
For $\bx_0$ a maximum,  
Theorem~\ref{thm:AOP91}(2) with $g =0$ (nonsingular case)
yields $N_\rsad \ge N_\rmax \ge 1$, which
yields $N \ge 3 $ (since $N_\rmin \ge 1$). 
For $\bx_0$ a saddle,  we get $N \ge 1 $ because there is also 
at least one minimum and the number of images must be odd. $\square$
\end{proof}

The sufficient condition $\kappa (\bx_0) > 1$ for multiple images of a source
at $\bby_0 = \bme (\bx_0)$ is not also a necessary condition.  A perturbed Plummer lens has
the following respective potential and surface mass density:
$$
\psi (\bx) = \frac{\kappa_0}{2} \log (1 + |\bx|^2) - \frac{\gamma}{2} (\ru^2 - \rv^2),
\qquad
\kappa (\bx) = \frac{\kappa_0}{(1 + |\bx|^2)^2},
$$
where $\gamma > 0$ is the external shear and $\bx = (\ru, \rv)$.
This lens is subcritical everywhere for $0 < \kappa_0 < 1$, but can still produce
multiple images (e.g., \cite[p. 429]{PLW}).

Multiple imaging is also discussed in Section~\ref{sec:mult-images-optgeom} from the optical geometry point of view, 
and for nonsingular lenses in a spacetime context in Section~\ref{sec:mult-images-spacetime}.

\subsection{Maximum Number of Images: Single Plane Lensing}
\label{sec:maximages-1p}

A natural next question is  to determine the maximum attainable number of
images due to $g$ stars.  Using a trick with complex quantities, Witt \cite{Witt90}
showed in 1990 that $g$ point masses will generate at most
$g^2 +1$ images.  Since it was unclear how to extend
the trick in \cite{Witt90} to multiplane lensing,
AOP \cite{Ptt97} gave in 1997 an alternative proof of the upper bound
using resultants, an approach generalizable to multiple lens planes
(see Section~\ref{sec:kplane-upperbound}).  

Combining the above upper bound result with Theorem~\ref{thm:AOP91}(2),
the number of images is bounded as follows:
\beq
\label{eq-lower-upper-N-a}
g + 1 \le N \le g^2 + 1.
\eeq
For two point masses $g =2$, equation (\ref{eq-lower-upper-N-a}) yields
$N = 3, 4$ or $5$. However, we cannot have $N = 4$ since by Theorem~\ref{thm:AOP91}(1)
an even number of stars has to produce an odd number of images.
Hence, 
$N = 3$ or $5$, which recovers the result in \cite{Sch-Weiss86}. 

It remained unclear whether the quadratic upper bound
of $g^2 +1$ in (\ref{eq-lower-upper-N-a}) can be attained. 
In 1997, Mao, AOP, and Witt \cite{MPW97} conjectured
that the maximum number of lensed images should
be linear in $g$. They also constructed  a lens system consisting
of $g$ point masses of equal mass $1/g$ on the vertices
of a regular polygon and showed that this symmetrical
system produces a maximum number of
$3g + 1$ images. The latter is a linear lower bound on
the maximum number of images for $g$ point masses.
Rhie \cite{Rhie03} showed in 2003 that by putting a mass $m_\epsilon >0$
at the center of the regular polygon, but with equal masses $1/(g-1)$ on its
vertices, a total of $5g-5$ images can be achieved for
sufficiently small masses.  Using a modification of
the method in \cite{MPW97}, Bayer and Dyer \cite{BD07}
gave in 2007 a much simpler proof of the result in \cite{Rhie03} and
improved our understanding of the result by 
determining an  upper bound $m_\star$ of the central mass such that the
maximum number $5g-5$ of images is attained for all $m_\epsilon <  m_\star$.
The conjecture was finally settled in 2006
when Khavinson and Neumann \cite{KN06} employed complex
rational harmonic functions to show that the total number
of images is at most $5g-5$.  The results are summarized below:
 
\begin{theorem}[Maximum Number of Images: Single-Plane Case] For $g\ge 2$ point masses on 
a lens plane and a light source not on a caustic, the
number of images satisfies:
  \begin{enumerate}
    \item  {\rm \cite{KN06}}   $N \le 5g -5$.
    \item  The upper bound $5g -5$ is attainable for: 
        \begin{enumerate}
 \item {\rm \cite{MPW97}} $g = 3$ if the point masses have equal mass $1/g$ and lie
on the vertices of an equilateral triangle centered at the origin
and inscribed in a circle of radius $r$ bounded above as follows:
$$
r < r_{\rm cr}  \equiv \left[\left(\frac{g-2}{g}\right)^{g-2/2}
 - \left(\frac{g-2}{g}\right)^{g/2}\right]^{g-2/g}. 
$$ 
\item {\rm \cite{Rhie03,BD07}} $g \ge 4$ if  $g-1$ of
the point masses have equal mass $1/(g-1)$ and lie
on the vertices of a regular polygon centered at the origin
and inscribed in a circle of radius $r < r_{\rm cr}$, and
a point mass of $m_\epsilon$ is at the center of the polygon
with mass upper bound given by:
$$
0< m_\epsilon <  m_\star
 \equiv r^{2 \left(\frac{g + 5}{g -1} \right)}  - \frac{r^6}{1 + r^6}.
$$ 
    \end{enumerate}
   \end{enumerate}
\end{theorem}
\smallskip

\noindent
We now have the following sharp (i.e., attainable) bounds on the total number of images
due to point masses:
\beq
\label{eq-lower-upper-N-b}
g + 1 \le N \le 5g -5, \hspace{0.5in} g \ge 2.
\eeq

\noindent
{\bf Remark:} For lensing by a general matter distribution, there is no overarching maximum number
of images because a mass clump can always be added to a lens
to create more images.

\section{Multiple Imaging in Multiplane Lensing}
\label{sec:IC-kplanes}

To set up the image counting results for multiple lens planes, we review some of the notation
and concepts needed. Consult \cite[Chap. 6]{PLW} for more details.

\begin{figure}
\centering
\includegraphics[width=0.7\textwidth, height=0.12\textheight]{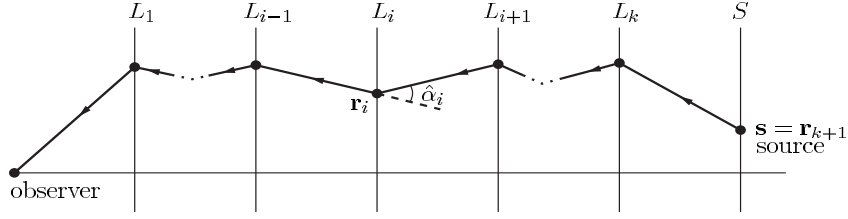}
\caption{A schematic of $k$-plane gravitational lensing.
The action of the lensing map is captured by
tracing light rays backwards from a subset $P$ of $L_1$
to the light source plane $S$. 
Credits: After \cite[p. 199]{PLW}.}
\label{fig:kplane-lensing}
\end{figure}

Let $L_i$ be the $i$th lens plane counting from observer to
the light source plane and set $\bX = (\bx_1, \dots, \bx_k) \in L_1 \times \cdots \times L_k$;
see Figure~\ref{fig:kplane-lensing}.
Denote the gravitational lens potential on $L_i$ by $\psi_i$, where
$i = 1, \dots, k$, and the light 
source plane
by $S = \RR^2$ with its elements
$\bby \in S$. 
Let $T_\bby^{(k)}: L_1 \times \cdots \times L_k \rightarrow \RR$,
be the $k$-plane time delay function induced by the potentials $\psi_i$ and 
denote the associated $k$-plane lensing map by $\bme^{(k)}: P \rightarrow S$.
Here $P = \RR^2 - B \subseteq L_1$, where  $B$ is the set of light path obstruction points.
Note that for single-plane lensing $P = L_1 = L$ and $B = A$ (set of singularities of
the potential $\psi_1 = \psi$).

The set $\Crit (\bme^{(k)})$ of {\it critical points} of $\bme^{(k)}$ is the locus of all
infinitely magnified lensed images for all light source positions in $S$, while
the set of {\it caustics} of $\bme^{(k)}$ is 
$ \Caustic (\bme^{(k)}) = \bme^{(k)} [\Crit (\bme^{(k)})]\subset S$, i.e., the set of all light source positions
from which there is at least one infinitely magnified lensed image. 
When $\Crit (\bme^{(k)})$ is bounded, then both $\Crit (\bme^{(k)})$ and
$\Caustic (\bme^{(k)})$ are compact \cite[p. 293]{PLW}.

The $k$-plane time delay function 
$T_\bby^{(k)}$
is called  {\it subcritical at infinity} if for each non-caustic point $\bby$ and
$|\bX|$ sufficiently large, the
eigenvalues of $\Hess T_\bby^{(k)} (\bX)$ are positive, i.e., the time delay surface  has
positive Gauss-Kronecker curvature for $|\bX|$ sufficiently large.
If $T_\bby^{(k)}$ is subcritical at infinity and 
$T_\bby^{(k)} (\bX) \rightarrow \infty$ as $|\bX| \rightarrow \infty$, then
$T_\bby^{(k)}$ is called {\it isolated}.

By Fermat's principle, the critical points of $T_\bby^{(k)}$ determine the 
lensed images of a source at $\bby$ (e.g., \cite{PLW}, Chaps. 3,6), so the lensed images are then identified
as generalized saddles of index $i$,
where $i = 0, 1, \dots, 2k$.
\\

\noindent
{\bf Notation:}
    \begin{enumerate}
 \item $N_i =$ number of images of index $i$. Here $N_0$ and
  $N_{2k}$ are, resp., the number of minima and maxima.
  \item $N_+ = \sum_{i {\rm (even)}} N_i =$ number of even index
       images.
  \item $N_- = \sum_{i {\rm (odd)}} N_i =$ number of odd index
       images.
    \item $N = N_+ + N_- = $ total number of images.
   \end{enumerate}

\subsection{Counting Formulas and Lower Bounds in Multiplane Lensing}
\label{sec:IC-kplanes-1}

In 1991, AOP \cite{Ptt91} 
extended to $k$-lens planes the image counting results of Theorem~\ref{thm:AOP91},
including the Odd Number Image Theorem.
All the results are topological in nature,
which give them wide applicability.
  As before, we state the results only for
the subcritical case (see \cite[Chap.12]{PLW} for more):

\begin{theorem}{\rm \cite{Ptt91}}  
\label{thm:AOP-kplane}
{\bf (Multiplane Counting Formulas and Lower Bounds)}
Let $T_\bby^{(k)}: L_1 \times \cdots \times L_k \rightarrow \RR$
be a  $k$-plane
time delay function
induced by lens potentials $\psi_1, \dots, \psi_k$, where each $\psi_i$ has
$g_i$ singularities, each of which is an infinite singularity.
Assume that $\bby$ is a non-caustic point and 
$T_\bby^{(k)}$ is isolated.
Then for a generic\footnote{Theorem holds either for $T_\bby^{(k)}$ or 
$T_\bby^{(k)} (\bX) + \hat{\bp} \cdot \bX$, for sufficiently small $\hat{\bp} \in \RR^{2k}$,
      except in a set of measure zero  \cite[pp. 452-455]{PLW}.}
$T_\bby^{(k)}$ the number of lensed images satisfies: 
\begin{enumerate}
 \item $\displaystyle N = 2 N_+ - \prod_{i = 1}^k (1-g_i) = 
     2 N_-  + \prod_{i = 1}^k (1-g_i)$, \qquad 
     $\displaystyle N_+ = N_- + \prod_{i = 1}^k (1-g_i)$.

 \item $\displaystyle N \ge \prod_{i = 1}^k (1 + g_i)$. 
  \item $N_0 \ge 1$, \quad
$\displaystyle N_j   \ge \sum_{ 1 \le \ell_1 < \cdots < \ell_j \le k} \ g_{\ell_1} \cdots
           g_{\ell_j}$ \ for \ $1 \le j \le k$,
\quad and \ $N_j \ge 0$ \ for \ $k+1 \le j \le 2k-1$.
\\
 \item If the corresponding lensing map $\bme^{(k)}$ is locally stable
and  $\Crit (\bme^{(k)})$ is bounded, then for sufficiently large $|\bby|$ the 
above lower bounds on
    the number of images of different types are attained:\\
       $\displaystyle N = \prod_{i = 1}^k (1 + g_i)$, 
\qquad
$N_0 = 1$, \qquad
$\displaystyle N_j   = \sum_{ 1 \le \ell_1 < \cdots < \ell_j \le k} \ g_{\ell_1} \cdots
           g_{\ell_j}$ \ for \ $1 \le j \le k$,\\
and 
\quad $N_j = 0$ \ for \ $k+1 \le j \le 2k-1$.
\end{enumerate}
\end{theorem}

%The Odd Number Image Theorem of Corollary~\ref{cor:AOP91} generalizes as follows to $k$ lens
%planes:

\begin{corollary}
\label{cor:AOP-kplane} 
{\bf (Multiplane Odd Number Image Theorem)}
For a non-caustic point $\bby$, let $T_\bby^{(k)}$ be a isolated $k$-plane  time delay function 
induced by nonsingular gravitational lens potentials. Then:\\
$$
N = 2 N_+ - 1 = 2 N_- + 1.
$$
\end{corollary}

\subsection{An Upper Bound on the Number of Images}
\label{sec:kplane-upperbound}

A natural question is whether there are multiplane lensing
bounds on the number of images analogous to the
single-plane ones in (\ref{eq-lower-upper-N-a}) or 
(\ref{eq-lower-upper-N-b}). The following 1997 theorem of AOP \cite{Ptt97}
is what is known so far:

\begin{theorem}{\rm \cite{Ptt97}}  
\label{thm:AOP97} 
{\bf (Multiplane Upper Bound)}
Let $\bby$ be a non-caustic point. Then the total number of
images due to 
gravitational lensing by $g$ point masses on $g$ lens planes with
one point mass on each lens plane is bounded as follows:
\beq
\label{eq-lower-upper-N-multiplane}
 2^g \le N \le 2 \left(2^{2(g-1)} - 1 \right), \qquad g \ge 2.
\eeq

\end{theorem}

By Theorem~\ref{thm:AOP-kplane}(1,4), 
the lower bound of $2^g$ is sharp and
the total number of images is always even. 
Theorem~\ref{thm:AOP97} was proven using the theory of resultants.

\section{General Stochastic Lensing: The Expected Number of Minimum Images}
\label{sec:stoch-lensing}

Stochastic lensing occurs when a component of a systems is random, typically,
the lens.
This relates to the broader study of random functions, a subject
that has been explored in mathematics primarily for Gaussian random fields (e.g.,
Adler and Taylor \cite{Adlerjon}, Azais and   Wschebor \cite{ds1}, Forrester and  Honner \cite{Forrester}, 
Li and Wei \cite{Li}, Shub and  Smale \cite{Smale}, Sodin and  Tsirelson
\cite{Sodin}, and references therein).
However, in gravitational lensing most of the realistic lensing scenarios  produce non-Gaussian random fields that have not been previously considered
(e.g., see Theorem~\ref{thm:pdfs-td-lm-sh} below).
Therefore,  a new mathematical  framework needs to be developed for the study of stochastic lensing. 

A natural first step in stochastic lensing is to study
the expectation of the random number of lensed images. We shall
present some recent rigorous mathematical work in that direction.

The expectation of $N_+ (D,{\bf y})$, the  number of positive parity lensed images inside a closed disk $D$ of a light source at position ${\bf y}$, is given by the following Kac-Rice 
type formula (see \cite{PRT2}):
%
%\begin{small}
\beq
\label{ricevar1}
 E[N_+ (D,{\bf y})] = \int_{D} E\left[ \left( (1-\kappa({\bf x}))^2 -  \Gamma^2({\bf x}) \right)  \ \one_{{\cal G}_A}({\bf x}) \,
\Big|  \ \boldsymbol{\eta} ({\bf x}) =  {\bf y}\right] \,         f_{\boldsymbol{\eta} ({\bf x})}  ( {\bf y} )   \ \D {\bf x},
\eeq
%\end{small}
%
%
for almost all ${\bf y}$. 
Here,  $\one_{{\cal G}_A}$ is the indicator function on
${\cal G}_A=\{\nu\in\R^2:G(\nu)\in (0,\infty)\}$, where
$ G({\bf x})=\det [ \Jac\bme ]({\bf x})$, and
$f_{\boldsymbol{\eta} ({\bf x})}$ 
is the probability density function (p.d.f.) of the lensing map at $\bx$.

This formula holds for a fixed light source position. Unfortunately,  this position is  unknown in most  lensing observations. Therefore, a physically relevant extension of (\ref{ricevar1}) is to view the light source position as a   random variable   whose probability density function is  compactly supported over a subset of the light source plane having positive measure.  This result can be further generalized to  the entire light source plane. To do so,  a countable compact covering  $\{\frkS\}$ of  light source plane is considered, with each set $\frkS$ in the family having the same positive area.  A corresponding family $\{Y_{\frkS}\}$ of  random light source positions Y is constructed, with each Y uniformly distributed over a set  $\frkS$. The {\it  global expectation} of the number of positive parity lensed images, denoted by $\widehat{E}[N_{+}(D,Y;{\mathfrak S})]_{\{{\mathfrak S}\}}$,  is then defined as the average of $ E[N_+ (D,y)] $ over the family $\{Y_{\frkS}\}$ (equivalently, over the family  $\frkS$). 
This notion was introduced in \cite{PRT2}.

AOP, Rider, and Teguia 2009 \cite{PRT2} determined a general formula for 
the global expected number of positive parity images:
\begin{theorem}\label{thm:globalmean00}{\rm \cite{PRT2}}
{\rm \bf (Global Expected Number of Positive Parity Images)}
The global expectation of the number of positive parity lensed images in $D$ is given by:
\beq
\label{globalmean00}
 \widehat{E}[N_{+}(D,Y;{\mathfrak S})]_{\{{\mathfrak S}\}}
=
\frac{1}{|{\mathfrak S}_0|} \ \int_{D} E\left[ \det \left[
\Jac\boldsymbol{\eta} \right]({\bf x}) \, \one_{{\cal G}_A}({\bf x})\right]  \ \D {\bf x},
\eeq
where $|{\mathfrak S}_0|=\area{({\mathfrak S})}$.
\end{theorem}
The importance of Theorem~\ref{thm:globalmean00} is its applicability to a wide range of lensing scenarios, with few assumptions on the  distribution of the random gravitational field within each scenario. Below we will discuss an application of this theorem to image counting in microlensing.

\section{Stochastic Microlensing: Asymptotics}
\label{sec:stoch-microlensing-asymp}

The mathematical analysis of the p.d.f.s and expectations in
microlensing is very difficult. An asymptotic approach is a natural
first step that will be important for understanding first order terms
and how deviations from them occur at subsequent orders.
For example, we shall see that, though the lensing map of microlensing is
a bivariate Gaussian at first order, the mapping deviates from
Gaussianity at the next orders. The p.d.f.s of the time delay function
and shear will also be seen to be non-Gaussian. Unfortunately, this
means that most of the technology already developed in the mathematical
theory of random fields will not be applicable directly to stochastic lensing. 

%\subsection{Global Expected Number of Minima in Microlensing Case}
%\label{sec:6.2}

\subsection{Notation}

The stochastic microlensing scenario we shall discuss is one with
uniformly distributed random star positions. 
 Recall from the Introduction that the
potential is given by
\[
 \psi_g(\bx)=\frac{\kappa_c}{2} |\bx|^2 -\frac{\gamma}{2}(\ru^2 - \rv^2)
       + \sum_{j=1}^g m_j  \log |\bx - \bxi_j |,
\]
where $\bx = (\ru, \rv)$,
the time delay function  at $\bby$ by
\[
     T_{g,\bby}(\bx) =   d_1 (\bx; \bby)
       - \sum_{j=1}^g m_j  \log | \bx - \bxi_j |,
\]
where
$d_1(\bx; \bby)= \frac{1}{2} |\bx - \bby |^2-\frac{\kappa_c}{2} | \bx |^2 +\frac{\gamma}{2}(\ru^2-\rv^2)$, 
and the components of the lensing map $\bme_g = (\bme_{1,g}, \bme_{2,g})$ by:
\beqan
\bme_{1,g}(\bx) 
&=& (1 - \kappa_c + \gamma) \ru
- \sum_{j=1}^g m \frac{U_j - \ru}{(U_j - \ru)^2 + (V_j - \rv)^2},
\\
\bme_{1,g}(\bx) 
&=& (1 - \kappa_c - \gamma) \rv
- \sum_{j=1}^g m \frac{V_j - \rv}{(U_j - \ru)^2 + (V_j - \rv)^2},
\eeqan
where $\bxi_j = (U_j, V_j).$
\smallskip

\noindent
{\bf Notation and Assumptions:}
\begin{enumerate} 
  \item Equal masses: $m_j=m,$ where $j=1,\cdots g$.
  \item $R = \sqrt{g/\pi}$ and $\kappa_* = \pi\, m$.
  \item $ B({\bf 0},R)$:  closed disc of radius $R$ centered at the origin $\bmo$.
  \item The random point mass positions $\xi_1, \dots, \xi_g$
are independent  and uniformly distributed
 over $ B({\bf 0},R)$.
\end{enumerate}

Now, normalize the random time delay function and random lensing map as follows:
$$
T_{g,\bby}^{*}(\bx)\equiv T_{g,\bby}(\bx)+ g m\, \log R,
\qquad \bme_g^* (\bx) \equiv \frac{\bme_g (\bx)}{\sqrt{\log g}}.
$$
Write the components of the random shear tensor due only to stars as follows:
$$
\Gamma_{1,g} (\bx) = \sum_{j=1}^g 
\frac{m[(U_j - \ru)^2 - (V_j - \rv)^2]}{[U_j - \ru)^2 + (V_j - \rv)^2]^2},
\qquad 
\Gamma_{2,g} (\bx) = \sum_{j=1}^g 
\frac{2 m (U_j - \ru)(V_j - \rv)}{[U_j - \ru)^2 + (V_j - \rv)^2]^2}.
$$
For fixed $\bby$ and $\bx$, we denote the possible values of the previous
random quantities as follows: 

\begin{enumerate}
 \item $\ttt =$ possible values of $T_{g,\bby}^{*}(\bx)$.
 \item $(\tth,\ttk)=$ possible values of $\bme_g^*(\bx)$.
 \item $(\ttz, \ttw)=$ possible values of $(\Gamma_{1,g}(\bx), \Gamma_{2,g}(\bx))$.
 \item $\ttG = (\ttz^2 + \ttw^2)^{1/2}=$  possible values of
                     the magnitude of the shear.
\end{enumerate}

For the asymptotic p.d.f. of the normalized random lensing map, we shall need
the quantities
$$
\tta_1 = \frac{(1-\kappa_c+\gamma)\ru}{\sqrt{\log g}}, \qquad
\tta_2 = \frac{(1-\kappa_c-\gamma)\rv}{\sqrt{\log g}}, \qquad
\sigma_g^* = \frac{\kappa_*}{\sqrt{\pi}} \, \sqrt{ \frac{\log(\ttB\,g^{1/2})}{\log g} }, 
\qquad
 \ttB =  \frac{2\sqrt{\pi}e^{1-\gamma_e}}{\kappa_*},
$$
where $\gamma_e$ is Euler's constant. For 
the asymptotic p.d.f. of the random shear, we shall use:
\beq
\label{eq:H1}
H_1(\ttG)  =  \kappa_*^2\,\frac{9\ttG^2 - 6\kappa_*^2}{4(\kappa_*^2 + \ttG^2)^{2}}
\eeq
and
\beq
 \label{eq:H2} 
H_2(\ttG;\, |\bx|)  =
\frac{\kappa_*|\bx|^2}{m}\frac{\kappa_*^2(6\kappa_*^2\, - \,
9\ttG^2 )}{2(\kappa_*^2 +\ttG^2)^{2}} - \frac{ \kappa_*^2(8
\kappa_*^4 - 24\kappa_*^2 \ttG^2 + 3 \ttG^4)}{4(\kappa_*^2 +
\ttG^2)^3}
+ \frac{15 \kappa_*^4 \left(8 \kappa_*^4 - 40 \kappa_*^2\ttG^2 + 15 \ttG^4\right)}{32 (\kappa_*^2 +  \ttG^2)^4}.
\eeq
Note that $H_2$ depends on $\kappa_* |\bx|^2/m$, which is the 
mean number of point masses within the
disc of radius $|\bx|$ centered at the origin.

\subsection{Asymptotic P.D.F.s of Random Time Delay Function, Random Lensing Map,
and Random Shear}
\label{sec:6.2.1}

In 2009,  AOP, Rider, and Teguia \cite{PRT1,PRT2} used a rigorous mathematical approach
to characterize up to order three
the asymptotic p.d.f.s of the microlensing normalized random time delay function, normalized random
lensing map, and random shear:

\begin{theorem} \label{thm:pdfs-td-lm-sh}
Fix $\bby \in S$ and $\bx = (\ru, \rv)\in B (\bmo, R)$.  Then:
\begin{enumerate}
 \item  {\rm \cite{PRT1}} {\bf (Random Normalized Time Delay Function)}  In the large $g$ limit,  the asymptotic p.d.f. of $T_{g,y}^{*}(x)$ 
takes the following form:
\beqan
\label{pdf-TDstar}
\sff_{T_{g,y}^{*}(\bx)}({\ttt})&=&\left\{
\begin{array}{lr}
\Bigl(\frac{2}{m}\Bigr)^g \frac{(\ttt-d_1-c)^{g-1}}{(g-1)!}\exp\left[-\frac{2(\ttt-d_1-c)}{m}\right], 
 &  \hspace{0.2in} \ttt >d_1+c\\
0, &  \ttt< d_1+c
\end{array}
\right\} \quad + \quad O(g^{-3/2}).
\eeqan
The first term  of the p.d.f.
is a Gamma distribution. 
\item {\rm \cite{PRT1}} {\bf (Random Normalized Lensing Map)}
In the large $g$ limit,  the asymptotic p.d.f. of  $\bme_g^*(\bx)$
takes the following form:
\beqa
\label{resultthm-pdf-lm}
\sff_{\bme_g^*(\bx)}(\tth,\ttk) &=&\frac{e^{-\frac{(\tth- \tta_1)^2+ (\ttk-\tta_2)^2}
 {2 (\sigma^*_g)^2}}}{\Bigl(\sqrt{2\pi}\sigma^*_g\Bigr)^2}
\left[1-\kappa_*\frac{(\tth- \tta_1)\ru +  (\ttk- \tta_2)\rv} {\sigma_g^2}\right. \nonumber\\
&& \hspace{1in}\left.  + \frac{\kappa_*^2}{4 \pi}\
\frac{\Bigl((\tth-\tta_1)^2+(\ttk-\tta_2)^2-2(\sigma^*_g)^2\Bigr)}{(\sigma^*_g)^4}
\log(\log g)\right]
\ + \ O\Bigl(\frac{1}{\log^2 g}\Bigr).
\eeqa
The first term of the p.d.f. 
is a bivariate Gaussian distribution, but  the next two terms highlight that
$\bme_g^*(\bx)$ already becomes non-Gaussian for large finite $g$.
\item {\rm \cite{PRT2}} {\bf (Random Shear)} In the large $g$ limit,  the asymptotic p.d.f. of  $(\Gamma_{1,g}^*(\bx), \Gamma_{2,g}^*(\bx))$
takes the following form:
\beq
\label{eq:pdf-shear}
\sff_{\Gamma_{1,g}(\bx),\Gamma_{2,g}(\bx)}(\ttz,\ttw)=
\frac{\kappa_*}{2\pi(\kappa_{*}^2 + \ttG^2)^{3/2}} \left[\ 1 + \ \frac{H_1(\ttG)}{g}  
\ + \ \frac{H_2(\ttG;\,|x|)}{g^2}\right] \, + \, O(g^{-3}),
\eeq
where $\ttG = (\ttz^2 + \ttw^2)^{1/2}$ denotes the possible values of the
magnitude of the shear.
The first term of the p.d.f. 
is a stretched bivariate Cauchy distribution. 
\end{enumerate}
\end{theorem}

By equation (\ref{eq:pdf-shear}), the asymptotic p.d.f. of the magnitude of the shear, namely,
$\Gamma_g(\bx) = \sqrt{\Gamma_{1,g}^2(\bx) + \Gamma_{2,g}^2(\bx)}$, 
is given as follows \cite{PRT2}:
$$
\sff_{\Gamma_g(\bx)}(\ttG)=\frac{\kappa_* \ttG}{(\kappa_*^2 + \ttG^2)^{3/2}} 
\left[1\, + \, \frac{H_1(\ttG)}{g}\ + \
\frac{H_2(\ttG; \,|x|)}{g^2}\right] + O(g^{-3}).
$$
\smallskip

\noindent
{\bf Remarks:}
\begin{enumerate}
 \item The first term in (\ref{resultthm-pdf-lm}) was basically found in 1986 by Katz, Balbus,
 and Paczy\'nski \cite{KB86}, who actually determined the
first term of the p.d.f. of the bending angle due only to stars, namely, the p.d.f. of the random
vector
$\bma_g (\bx) = \bme_g (\bx) - ((1 - \kappa_c + \gamma) \ru, (1 - \kappa_c - \gamma) \rv)$.
 \item The first term in (\ref{eq:pdf-shear}) was found in 1984 by Nityananda and Ostriker \cite{NO84}. 
\end{enumerate}

\noindent
Subsequent work in 2009 by Keeton \cite{Ke09} used 
semi-analytical and numerical methods to study the
stochastic properties of the lens potential, deflection angle, and shear under
different assumptions about the distributions of the stars' masses and
positions.

\subsection{Global Expected Number of Micro-Minima}

Wambsganss, Witt, and Schneider 1992 \cite{WWS}
determined the
limit $g \rightarrow \infty$ of the 
(global\footnote{The terminology ``global expected number of minima''
    was actually introduced in \cite{PRT2}, where the difference between
    the expected number of minima and global expected number of minima was
    clarified.})
expected number $\ttE_0$ of minima in the {\it entire} plane $\RR^2$
for microlensing without shear.
This was extended to the case with shear in
2003 by Granot, Schechter, and Wambsganss \cite{GSW}:
$$
%\widehat{E}[N_{g,min}(\RR^2, Y; {\mathfrak
%S})]_{\{{\mathfrak S}\}}
%=
\ttE_0  =
\frac{\kappa_*}{2 \pi|(1 - \kappa_{\rm tot})^2 - \gamma^2|}
\int_{{\cal B}} \frac{(1-\kappa_c)^2 - (\gamma+ \ttz)^2 -
\ttw^2  }{(\kappa_{*}^2 + \ttG^2)^{3/2}} \rd \ttz \rd \ttw,
$$ 
where $\kappa_{\rm tot} = \kappa_* + \kappa_c$ and $\gamma$ 
is same shear employed at the microlensing scale.

The value $\ttE_0$ resulting from the limit $g \rightarrow \infty$ is, of course,
independent of $g$ and can be treated as the first term 
in an asymptotic expansion in $1/g$. 
On the other hand, since the first term $\ttE_0$ is independent of $g$,
if we have only the term $\ttE_0$, then
there is no way to know analytically the smallest $g$ needed
to lie
within a certain approximation of the  global expected number of micro-minima.
This is especially important in numerical simulations. 
Therefore, we need to find more terms in the asymptotic expansion.

AOP, Rider and Teguia 2009 \cite{PRT2}
applied  Theorem~\ref{thm:globalmean00}
and Theorem~\ref{thm:pdfs-td-lm-sh}(3)
to determine rigorously the global expected number of micro-minima
up to three orders:

\begin{theorem} \label{thm:gexp-micro}
{\rm \cite{PRT2}}
{\rm \bf (Global Expected Number of Micro-Minima)}
Let $D$ be a closed disc and suppose that
continuous matter is subcritical, i.e., $0 \leq \kappa_c <1$. Then,
in the large $g$ limit, the first three asymptotic terms  of the
global expectation of the number of  minimum images in
$D$  is given by:\footnote{The quantities $H_1$ and $H_2$ were defined 
  in  (\ref{eq:H1}) and (\ref{eq:H2}).}
%
%
%\begin{small}
\beqa 
\label{eq:gem-micro}
 \widehat{E}[N_{g,min}(D, Y; {\mathfrak
S})]_{\{{\mathfrak S}\}}
 &=&\frac{\kappa_* \ \mu_{D,{\mathfrak
S}_0}}{2\pi} \int_{{\cal B}} \frac{(1-\kappa_c)^2 - (\gamma+ \ttz)^2 -
\ttw^2  }{(\kappa_{*}^2 + \ttG^2)^{3/2}} \left[ 1   +
\frac{H_1(\ttG) }{g}   +  \frac{H_2(\ttG; a_0) }{g^2}   \right] \D \ttz \D \ttw  +
O(g^{-3}),\nonumber \\
\eeqa
%\end{small}
%
where
$
\mu_{D,{\mathfrak S}_0}=\frac{|D|}{|{\mathfrak S}_0|}$,
${\cal B}$ is a closed disc 
of radius $1-\kappa_c$ centered at
$(-\gamma,0)$,
and 
$a_0 =  \frac{1}{|D|}\int_D |{\bf u}|^2 \D {\bf u}.
$
\end{theorem}

Theorem~\ref{thm:gexp-micro} gives us 
the leading three terms
 in an asymptotic expansion of
the global expected number of micro-minima in any  reference disc $D$,
not necessarily in the entire plane.
The first term in that expansion (\ref{eq:gem-micro})
is more general than $\ttE_0$ because 
$\ttE_0$ applies to the whole plane. 
A possible physical model for the factor $\mu_{D,{\mathfrak S}_0}$ 
in (\ref{eq:gem-micro}) is 
the macro-scale magnification given by:
$$
\mu_{D,{\mathfrak S}_0}=\frac{1}{|(1 - \kappa_{\rm tot})^2 - \gamma^2|}.
$$
It is only in this case that the first term in 
(\ref{eq:gem-micro})
 coincides with $\ttE_0$.

Another consequence of Theorem~\ref{thm:gexp-micro} is it
enables us to estimate how small we can choose $g$ in order to have the first
three terms in the asymptotic expansion (\ref{eq:gem-micro}) lie within a certain
percentage of the global expected number of micro-minima.  
Theorem~\ref{thm:gexp-micro} also shows us analytically 
how accurate an approximation
the first term is to the exact global expectation by quantifying
how  the next two higher-order terms perturb the
first one.
\smallskip

\noindent
{\bf Remark:}
The third-order term in 
 (\ref{eq:gem-micro})
depends on the mean number of point masses within the
disc of radius $|\bx|$ centered at the origin, namely, 
 the quantity $\kappa_* |\bx|^2/m$;  compare equation (\ref{eq:H2}).

\section{Multiple Images in Optical Geometry}
\label{sec:mult-images-optgeom}

\subsection{The Optical Metric and Fermat's Principle}
\label{sec:4.1}
The trajectories of spatial light rays can also be studied in optical geometry, which is conceptually between the thin-lens, weak-deflection approximation used in the previous sections and the full spacetime treatment of null geodesics. Optical geometry, which is also known as Fermat geometry or optical reference geometry, is a useful tool to investigate gravitational and inertial forces in General Relativity \cite{abram}. Recently, it has also been applied to field theory near black hole horizons by Gibbons and Warnick \cite{gibbons2}, and to analogue models of gravity in Finsler geometry by Gibbons, Herdeiro, Warnick and MCW \cite{gibbons5}.

Fermat's principle and optical geometry in the conformally stationary case was discussed in detail by Perlick \cite{perlick1b}. To illustrate this approach, consider for simplicity a static spacetime metric 
$$
\D s^2=g_{00}\, (\D t)^2 + g_{ij}\, \D x^i \D x^j,
$$ 
where the summation convention is used,
and a null curve parametrized by $0\leq \nu \leq 1$, say, with tangent vector $\sfk$. Then the variational principle yields (e.g., Frankel \cite{frankel}):
\[
\delta t=\frac{1}{g\left(\frac{\partial}{\partial t},\sfk\right)(1)}
\int_0^1 g(\delta x, \nabla_\sfk \sfk)\D \nu=0,
\label{fermatsprinciple}
\]
if the null curve is in fact a null geodesic with $\nabla_\sfk \sfk=0$. Hence, one obtains Fermat's principle of stationary arrival time, rather than stationary travel time as in the case of flat space used in the impulse approximation. Now recasting the spacetime line element as follows, we see that the spatial light rays are geodesics of the optical metric $g^{\mathrm{opt}}$ with Riemannian signature,
\[
(\D t)^2=-\frac{g_{ij}}{g_{00}}\, \D x^i \D x^j \equiv g^{\mathrm{opt}}_{ij} \, \D x^i \D x^j,
\label{opticalmetric}
\]
by Fermat's Principle.

\subsection{Gauss-Bonnet and Lensed Images}
\label{sec:4.2}
\begin{figure}
\centering
\includegraphics[width=0.5\textwidth]{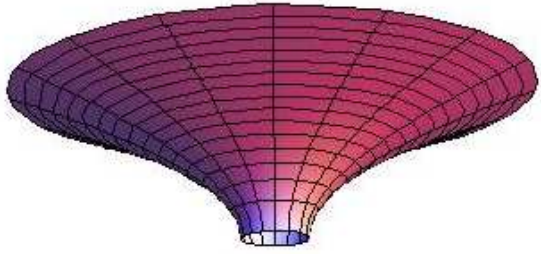}
\caption{Isometric embedding in $\R^3$ of the equatorial plane of the Schwarzschild solution in the optical metric. The waist occurs at the radius of the photon sphere, $r=3m$. The Gaussian curvature is negative everywhere.}
\label{fig:opticalmetric}
\end{figure}

Optical geometry offers a different perspective on image multiplicity in gravitational lensing, which is also partially topological. This can be understood with the Gauss-Bonnet Theorem, which connects the local optical geometry with global properties of the light rays. Gibbons \cite{gibbons}
discussed this for cosmic strings, and this method was recently extended to spherically symmetric metrics by Gibbons and MCW \cite{gibbons3} and Gibbons and Warnick \cite{gibbons2}. In this approach, the occurrence of two images can be expressed as a digon of light rays:
 
\begin{theorem}{\rm \cite{gibbons,gibbons3,gibbons2}}  {\rm \bf (Light Ray Digons)}
Let $(S,g^{\mathrm{opt}})$ be a totally geodesic, simply connected surface endowed with an optical metric $g^{\mathrm{opt}}$ and Gaussian curvature $K$. Let $D\subset S$ be a geodesic 
digon\footnote{A {\it geodesic digon} in a lensing scenario 
           is a polygon with two vertices, bounded by the two
           geodesics intersecting at the source and the observer.}
 bounded by two light rays intersecting at the light source $s\in S$ and the observer $o \in S$ with corresponding positive interior angles $\theta_s$ and $\theta_o$. Then:
\[
\theta_s+\theta_o=\int\int_D K \D S.
\]
\end{theorem}

\smallskip

\noindent
{\bf Example:}
The Schwarzschild solution with mass parameter $m$ is an instructive example. The equatorial plane in the optical metric outside the photon sphere is shown in Fig. \ref{fig:opticalmetric}, and it can be seen that the Gaussian curvature is negative everywhere. Indeed, a calculation of the Gaussian curvature gives
\[
K =-\frac{2m}{r^3\left(1-\frac{2m}{r}\right)^{3/2}}\left(1-\frac{3m}{2r}\right) < 0.
\]
Spatial light rays, which are the geodesics of $S$, must therefore diverge locally, and the equation of the Theorem cannot be fulfilled. So the fact that two light rays can intersect at the source $s$ and the observer $o$ in Schwarzschild geometry shows that $S$ cannot be simply connected, which indeed it is not because of the event horizon at $r=2m$. Hence, the topological contribution to the Gauss-Bonnet theorem turns out to be essential for image multiplicity.
\smallskip

\noindent
{\bf Example:}
Now consider the Plummer model with mass $m_0$ and scale radius $r_0$. This is a reasonable model for an extended non-relativistic gravitational lens (i.e., a galaxy) which allows multiple imaging: three images are produced if the light source is in the maximal caustic domain. Of course, there is no event horizon in this model, so the surface $S$ is simply connected, and the optical metric approaches that of the Schwarzschild solution with $K<0$ at large radii $r$, since $m_0$ is finite. Therefore, the equation of the theorem holds, and we expect that the Gaussian curvature of the optical metric changes sign in order to allow for multiple images. A calculation of $K$ to lowest order yields
\[
K=-\frac{2m_0}{r_0^3}
\left(1+\left(r/r_0\right)^2\right)^{-\frac{3}{2}}
\left(1-\frac{3}{1+\left(r/r_0\right)^2}\right)+  O (m_0^2),
\]
which confirms that  $K>0$ for small radii.

A more detailed analysis shows that the Gauss-Bonnet theorem can also be used to calculate deflection angles. In the case of the singular isothermal sphere, for instance, the optical metric describes a cone so that the gravitational deflection emerges as a consequence of the cone's deficit angle, rather like a spatial analogue of lensing by cosmic strings \cite{gibbons3}.

\section{Multiple Images in Spacetime}
\label{sec:mult-images-spacetime}

\subsection{Necessary and Sufficient Conditions for Multiple Images}
\label{sec:5.1}

The theories for spatial light rays provide useful frameworks for gravitational lensing, 
especially since the impulse approximation can easily be applied to models of astrophysical interest. However, the fundamental arena of optics in General Relativity is of course spacetime. The study of wavefront singularities in spacetime was pioneered by Friedrich and Stewart \cite{friedrich}, who classified stable singularities in Minkowski space and considered the relationship with the initial value problem in General Relativity. The general form of Fermat's Principle in spacetime was investigated by Kovner \cite{kovner}, and a precise proof for arbitrary Lorentzian manifolds $(M, g)$ was first obtained by Perlick \cite{perlick1} in 1990. This also laid the foundation for a rigorous study of the conditions for multiple images in spacetime, which made an earlier result by Padmanabhan and Subramanian \cite{paddy} more precise. Perlick showed that the existence of a conjugate point or a cut point along a null geodesic is a sufficient condition, which becomes necessary if further conditions on the topology or causal structure of $M$ are imposed:
\begin{theorem}{\rm \cite{perlick3}}
{\rm \bf (Multiple Images in Spacetime)}
%\newline
Let $(M,g)$ be a four-dimensional time-oriented Lorentzian manifold.
\begin{enumerate}
\item{{\bf Sufficient Conditions:}} Let $\lambda$ be a future-pointing null geodesic affinely parametrized by $s$, and fix $\lambda(s_1)=p \in M$.
\begin{enumerate}
 \item If there is a $s_2>s_1$ such that $\lambda(s_2)$ is conjugate to $\lambda(s_1)$ along $\lambda$, then there is a timelike curve through each $\lambda(s), s>s_2$, which can be reached from $p$ along another future-pointing null geodesic.
\item If there is a $s_2>s_1$ such that $\lambda(s_2)$ is the future cut point of $p$ along $\lambda$, then there is a timelike curve through each $\lambda(s), s>s_2$, which can be reached from $p$ along another future-pointing null geodesic.
\end{enumerate}
\item{{\bf Necessary Conditions:}} Fix a timelike curve $\gamma$ and a point $p \in M$.
\begin{enumerate}
\item If there are two future-pointing null geodesics from $p$ to $\gamma$ which are null homotopic, then there is a future-pointing null geodesic from $p$ to $\gamma$ which contains a point conjugate to $p$.
\item If $(M,g)$ is strongly causal and if there are two future-pointing null geodesics from $p$ to $\gamma$, then the intersection with $\gamma$ comes on or after the future cut point of $p$ along at least one of the null geodesics.
\end{enumerate}
\end{enumerate}
\end{theorem}
Having established conditions for existence, one can now proceed with counting results for null geodesics.

\subsection{The Odd Number Theorem in a Spacetime Setting}
\label{sec:5.2}
The proofs of the Odd Number Theorem for single and multiple lens planes discussed in 
Sections~\ref{sec:IC} and \ref{sec:IC-kplanes}  employ finite dimensional Morse theory in the impulse approximation. Interestingly, this result can also be extended to spacetime under certain assumptions. McKenzie 1985 showed this using the degree of a map between two spatial spheres, which requires stationarity of the spacetime, and gave a proof using Uhlenbeck's Morse theory for null geodesics on globally hyperbolic Lorentzian manifolds \cite{mckenzie}. However, these conditions appear to be too restrictive for realistic spacetimes, as discussed by Gottlieb 1994 \cite{gottlieb}. An extension to infinite dimensional Morse theory on the Hilbert space of null curves was developed 
Perlick 1995 \cite{perlick2} and Giannoni, Masiello and Piccione 1998 \cite{giannoni1}. The resulting Morse relations have been applied by Giannoni and Lombardi \cite{giannoni2} to prove a version of the Odd Number Theorem. In 2001, Perlick \cite{perlick4} defined the concept of a \textit{simple lensing neighborhood}, which formalizes a physically meaningful lensing geometry to avoid some of the technical complexities, and proved the following spacetime version of the Odd Number Theorem:

\begin{theorem}{\rm \cite{perlick4}}
{\rm \bf (Odd Number of Null Geodesics)}
 {\rm \cite{perlick4}}
Let $U$ be a simple lensing neighborhood in a four-dimensional time-oriented Lorentzian manifold $(M,g)$. Fix a point $p\in U$ and a timelike curve $\gamma$ in $U$ such that it has no endpoints on the boundary $\partial U$. If $\gamma$ intersects neither $p$ nor the caustic of the past light cone of $p$, then the number of past-pointing null geodesics from $p$ to $\gamma$ completely within $U$ is finite and odd.
\end{theorem}
A rigorous treatment of the optics in a spacetime setting can be found in Perlick \cite{perlick5}.

\section{Global Magnification Relations for Special Lens Models}
\label{sec:glob-magrel}

In gravitational lensing, when a source gives rise to multiple images, the magnifications of these images often obey certain relations.  The simplest example of such a relation is 
provided by a single point-mass lens.  In this setting a source gives rise to two lensed images, and it can be shown that their signed magnifications always sum to unity:
$$\mu_1+\mu_2 = 1\ \ (\mbox{one point mass})
$$
(e.g., \cite{PLW}, p. 191).  The surprising fact about this result is that it holds irrespectively of the mass of the point mass and its position on the lens plane.  It is also a ``global" relation, by which is meant that it involves all of the images of a given source and not merely a subset of them.  Witt and Mao 1995 \cite{Witt-Mao} generalized this result to a two point-mass lens.  They showed that for a source lying anywhere inside the caustic curve, a region which gives rise to five lensed images (the maximum number in this case), the sum of the signed magnifications of these images is also unity:
$$
\mu_1 + \mu_2 + \mu_3 + \mu_4 + \mu_5 = 1\ \ (\mbox{two point masses}).
$$
Like its predecessor, this relation is ``global" and holds irrespectively of the lens's configuration, provided the source lies inside the region giving rise to the maximum number of images.  Rhie 1997 \cite{Rhie} subsequently extended this result to $N$ point masses.  Dalal 1998 \cite{Dalal} and Witt and Mao 2000 \cite{Witt-Mao2} then showed that other common lens models, such as singular isothermal spheres and ellipses (SISs and SIEs) and elliptical power-law potentials, possess similar global magnification relations, even when these models include shear.

Dalal and Rabin 2001 \cite{Dalal-Rabin} provided a residue approach that systematized and expanded
on the above work.  They began with the Euler trace formula, which they proved using residue calculus.  The Euler trace formula identifies sums of magnifications as coefficients of certain coset polynomials; we shall return to this in Section~\ref{sec:proof-ADEthm} below.  The
method in \cite{Dalal-Rabin} uses meromorphic differential forms in several complex variables.  By considering a meromorphic 2-form on $\mathbb{C}^2$ consisting of polynomials whose common zeros are the image positions of the lensing map, the authors were able to use the Global Residue Theorem, which states that on compact manifolds the sum of all the residues of a meromorphic form vanishes.  This allows one to replace the common zeros of a form in $\mathbb{C}^2$, viewed as a subset of the compact manifold $\mathbb{CP}^2$, by minus the sum of residues at infinity in $\mathbb{CP}^2$.  In this way magnification sums were transformed to a condition about the behavior of the lens equation at infinity.  Their results are summarized in the following thereom:

\begin{theorem}{\rm \cite{Dalal-Rabin}}
\label{dalal:theorem}
The models listed below possess the following zeroth and first magnification moment relations:
\begin{center}
\begin{tabular}{ c  c  c }
\hline
 & &  \\
{\rm Model} & $\displaystyle\sum_j \mu_j$ & $\displaystyle\sum_j \mu_j z_j$ \\
 & & \\
\hline
& & \\
point masses & $1$ & $\displaystyle z_s + \sum_j \frac{m_j}{\overline{z}_s-\overline{z}_j}$\\
& & \\
point masses + shear & $\displaystyle \frac{1}{1-\gamma^2}$ & $\displaystyle \frac{z_s+\gamma \overline{z}_s}{(1-\gamma^2)^2}$\\
& & \\
SIE & $2$ & $2z_s$ \\
& & \\
SIE + elliptical potential & $1$ & $\displaystyle z_s + 2\gamma\overline{z}_s - 
\frac{\overline{z}_s^3}{32\, b^2 \, \gamma^2}$ \\
& & \\
SIS + shear & $\displaystyle \frac{2}{1-\gamma^2}$ & $\displaystyle \frac{2(z_s+\gamma \overline{z}_s)}{(1-\gamma^2)^2}$ \\
& & \\
SIE + shear & $\displaystyle \frac{2}{1-\gamma^2}$ & $\displaystyle \frac{2(z_s+\gamma e^{2i\theta_{\gamma}}\overline{z}_s)}{(1-\gamma^2)^2}$\\
& & \\
\hline
\end{tabular}
\end{center}
{\rm Notation: $m_j$ is the mass of $j$th point mass, $\gamma$ is the shear with orientation $\theta_{\gamma}$,
and $z_j$  and $z_s$ are, respectively, the position of the $j$th lensed image and
the position of the source using complex variables.}
\end{theorem}

Residue calculus methods were also used by Hunter and Evans 2001 \cite{Hunter-Evans},
wherein magnifications of images were realized as residues of complex integrands.  By Cauchy's theorem, sums of magnifications are then equivalent to a contour integral.  This method was used to derive magnification relations for elliptical power-law potentials which expanded upon the work of \cite{Witt-Mao2}.  As first shown in \cite{Witt-Mao2}, for an elliptical power law potential $\psi (\bx) \propto (\ru^2+\rv^2q^{-2})^{b/2}$,
where  $q$ is the ratio of the minor to major axes,
the total signed magnification denoted by ${\sf B}$  is exactly ${\sf B} = 2/(2-b)$ for the cases $b = 0,1$; for other values, it is an approximation.  
The contour integral method used in \cite{Hunter-Evans,Evans-Hunter} covered not only all cases when ${\sf B}$ is an integer, including second and third magnification moments, reciprocal moments, and random shear, but also cases with non-integer values of ${\sf B}$ as well. 
In 2002, Evans and Hunter \cite{Evans-Hunter} extended their results 
in \cite{Hunter-Evans} to include elliptical power-law potentials with a core radius, and calculated magnification invariants for subsets of the images with even and odd parities.

\section{Local Magnification Relations:  Caustics up to Codimension 3}
\label{sec:loc-magrel}

\subsection{Quantitative Fold and Cusp Magnification Relations: Single Plane Lensing}
\label{sec:quant-magrel}

All of the above magnification relations are ``global" because they involve all the lensed images of a given source.  But they are not {\it universal} because their relations were derived in the context of specific types of lens models (point-mass lenses, SIEs, etc.)  There is another type of magnification relation, a so-called ``local" magnification relation, that {\it is} universal in the sense that it holds for a generic family of lens models.  It is called a ``local" relation because it holds for a subset of the total number of lensed images.  Such relations arise when the source lies close to a caustic singularity.  The two simplest types of such singularities are the fold and the cusp.  For a source near a fold, there will be two images straddling the critical curve, while for a source near a cusp, there will be a triplet of images.  Interestingly, the signed magnifications of this doublet and triplet always sum to zero (e.g., Blandford and Narayan 1986 \cite{Blan-Nar}, Schneider and Weiss 1992 \cite{Sch-Weiss92}, and Zakharov 1999 \cite{Zakharov}:
\beqa
\label{cusp}
\mu_{1} + \mu_{2} = 0\ {\rm (fold)},\hspace{0.5 in}\mu_{1} + \mu_{2} + \mu_{3} = 0\ {\rm (cusp)}.
\eeqa
These relations are important in gravitational lensing because they can be used to detect dark substructure in galaxies with ``anomalous" flux ratios.  These anomalies arise as follows.  For quasars with four lensed images, it is often the case that the {\it smooth} mass densities used to model the galaxy lens reproduce the number and positions of the lensed images, but fail to reproduce the image flux ratios.  Mao and Schneider 1998 \cite{Mao-Sch} showed that in such situations the cusp magnification relation (\ref{cusp}) fails.  They attributed this failure to the assumption of smoothness in the galaxy lens, and argued that this smoothness breaks down on the scale of the image separation.  This suggests the presence of substructure in the galaxy lens.  The possibility of this became even more intriguing when Metcalf and Madau 2001 \cite{M-M} and Chiba 2002 \cite{Chiba} showed that dark matter would be a natural candidate for this substructure.

Being able to identify which anomalous lenses have substructure now became a priority.  
Keeton, Gaudi and AOP in 2003 \cite{KGP-cusps} and 2005 \cite{KGP-folds} 
developed a rigorous framework by which to do so, using the observable quantities:
$$
R_{\rm{fold}} \equiv {\mu_{\rm{1}} + \mu_{\rm{2}} \over |\mu_{\rm{1}}| +
|\mu_{\rm{2}}|} = {F_{\rm{1}} - F_{\rm{2}} \over F_{\rm{1}} +
F_{\rm{2}}}\ , \qquad
R_{\rm{cusp}} \equiv {\mu_{1} + \mu_{2} + \mu_{3} \over |\mu_{1}| +
|\mu_{2}| + |\mu_{3}|} = {F_{1} - F_{2} + F_{3} \over F_{1} + F_{2} +
F_{3}}\ ,
$$
where $F_i$ is the observable flux of image $i$.  The importance of these quantities is as follows.  If a source lies sufficiently close to a fold or cusp caustic, then (\ref{cusp}) predicts that $R_{\rm{fold}}$ and $R_{\rm{cusp}}$ should vanish.  If $R_{\rm{fold}}$ and $R_{\rm{cusp}}$ deviate significantly from zero (Monte Carlo methods were used to determine what constituted a significant deviation), then that would indicate the presence of substructure in that particular lens.  On this basis it was shown in \cite{KGP-cusps,KGP-folds} that for the multiply imaged
lens systems they analyzed, 5 of the 12 fold-image lenses 
and 3 of the 4 cusp-image lenses showed evidence of substructure.

%\subsection{Magnification Relations for Higher-Order Caustics: Algebraic Approach}
%\label{sec:8.2}

\subsection{Quantitative Elliptic and Hyperbolic Umbilics' Magnification Relations: 
Single Plane Lensing}
\label{sec:quant-hu-eu}

Consider a family of time delay functions $T_{c,\bby}$ which induces a corresponding family of lensing maps $\blm_{c}$.  Here $\bby$ is the source position on the source plane and the parameter 
$c$ can be any physical input, such as the core radius or external shear.\footnote{There is at most
  one universal unfolding parameter $c$ for caustics up to codimension $3$.}  Using rigid coordinate transformations and Taylor expansions, the {\it universal, quantitative} form of the lensing map can be derived in a neighborhood of a caustic (see \cite{SEF,PLW}).  As mentioned above, the quantitative forms of lensing maps near fold and cusp caustics obey the fold and cusp magnification relations (\ref{cusp}).   Aazami and AOP 2009 \cite{Aazami-Petters1} showed recently that the quantitative forms of lensing maps near certain {\it higher-order} caustic singularities, namely the elliptic and hyperbolic umbilics, also satisfy magnification relations analogous to (\ref{cusp}).  Their work is summarized by the following theorem:

\begin{theorem}{{\rm\cite{Aazami-Petters1}}}
\label{theorem:quant}
For any of the smooth generic family of time delay functions $T_{c,\bby}$ corresponding to the elliptic umbilic or hyperbolic umbilic caustic singularities, and for any source position in the indicated region, the following results hold:
\begin{enumerate}
\item $D_{4}^{-}$ {\rm (Elliptic Umbilic)} satisfies the following magnification relation in its four-image region: 
$$
\mu_{1} + \mu_{2} + \mu_3 + \mu_4 = 0.
$$
\item $D_{4}^{+}$ {\rm (Hyperbolic Umbilic)} satisfies the following magnification relation in its four-image region: $$\mu_{1} + \mu_{2} + \mu_3 + \mu_4 = 0.$$
\end{enumerate}
\end{theorem}

An application of this theorem to substructure studies was also given in \cite{Aazami-Petters1}, using the hyperbolic umbilic ($D_4^{+}$) in particular.  Analogous to the observables $R_{\rm{fold}}$ and $R_{\rm{cusp}}$, the authors considered the following quantity:
$$
R_{\rm{h.u.}} \equiv {\mu_{1} + \mu_{2} + \mu_{3} + \mu_{4} \over
|\mu_{1}| + |\mu_{2}| + |\mu_{3}| + |\mu_{4}|} = {F_{1} - F_{2} + F_{3} -
F_{4} \over F_{1} + F_{2} + F_{3} + F_{4}}\cdot \nonumber
$$
By Theorem~\ref{theorem:quant}, $R_{\rm{h.u.}}$ should vanish for a source lying sufficiently close to a hyperbolic umbilic caustic singularity and lying in the four-image region.  One advantage of $R_{\rm{h.u.}}$ is that it incorporates a larger number of images than $R_{\rm{fold}}$ and $R_{\rm{cusp}}$, and also applies to image configurations that cannot be classified as fold doublets or cusp triplets.  In fact recent work has shown that higher-order caustics like the hyperbolic umbilic can be exhibited by lens galaxies.  Evans and Witt 2001 \cite{Evans-Witt}, Shin and Evans 2007 \cite{Shin-Evans}, and Orban de Xivry and Marshall 2009 \cite{Or-Marshall} have shown that realistic lens models can exhibit swallowtail ($A_5$) and butterfly caustics ($A_4$), as well as elliptic umbilics ($D_4^{-}$) and hyperbolic umbilics ($D_4^{+}$).  It is hoped that such lensing effects will be seen by the Large Synoptic Survey Telescope, and thus that higher-order relations such as $R_{\rm{h.u.}}$ will become applicable in the near future.

\begin{figure}[tbh]
\label{cuspcross}
\begin{center}
\begin{tabular}{| c |}
\hline
\\
$~~~~~$\includegraphics[scale=.25]{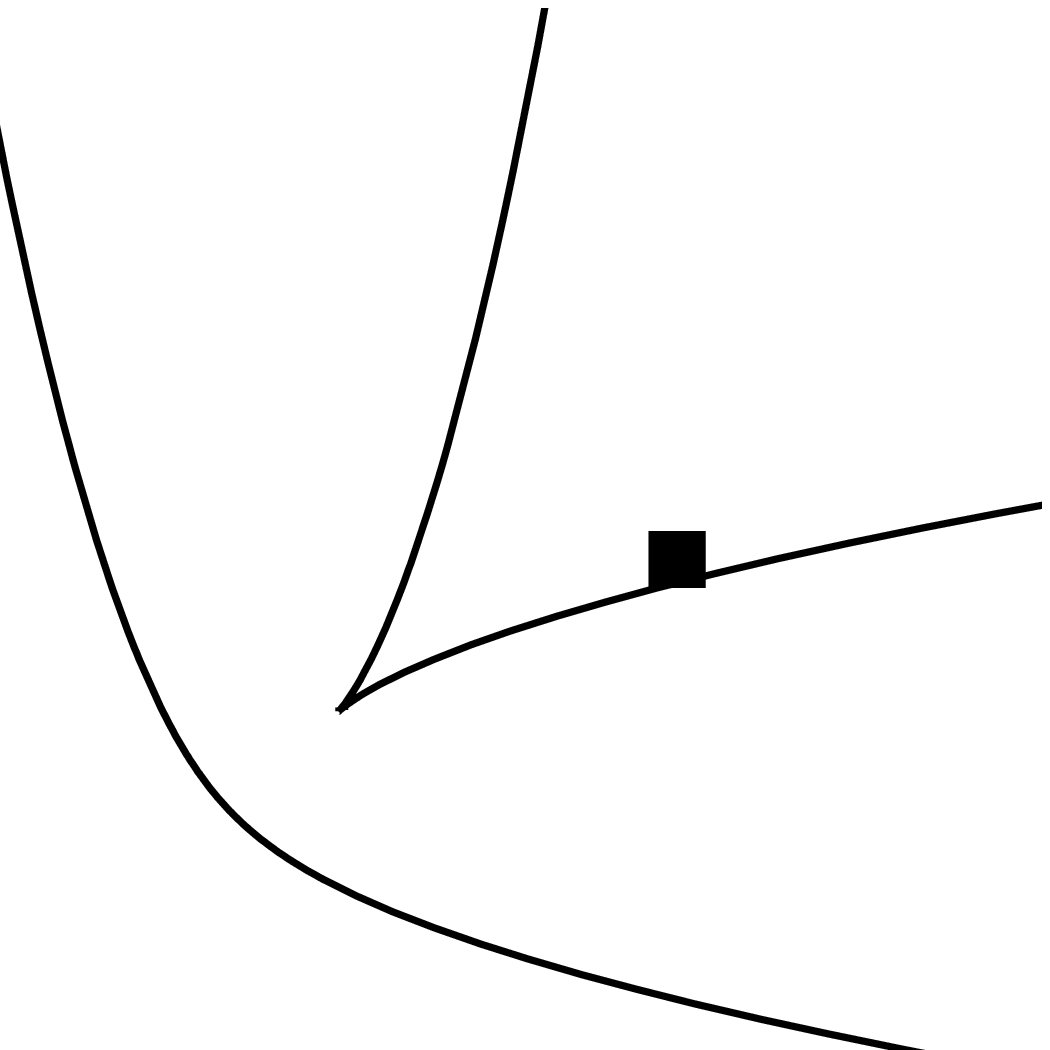}$~~~~~~~~~~$
\includegraphics[scale=.25]{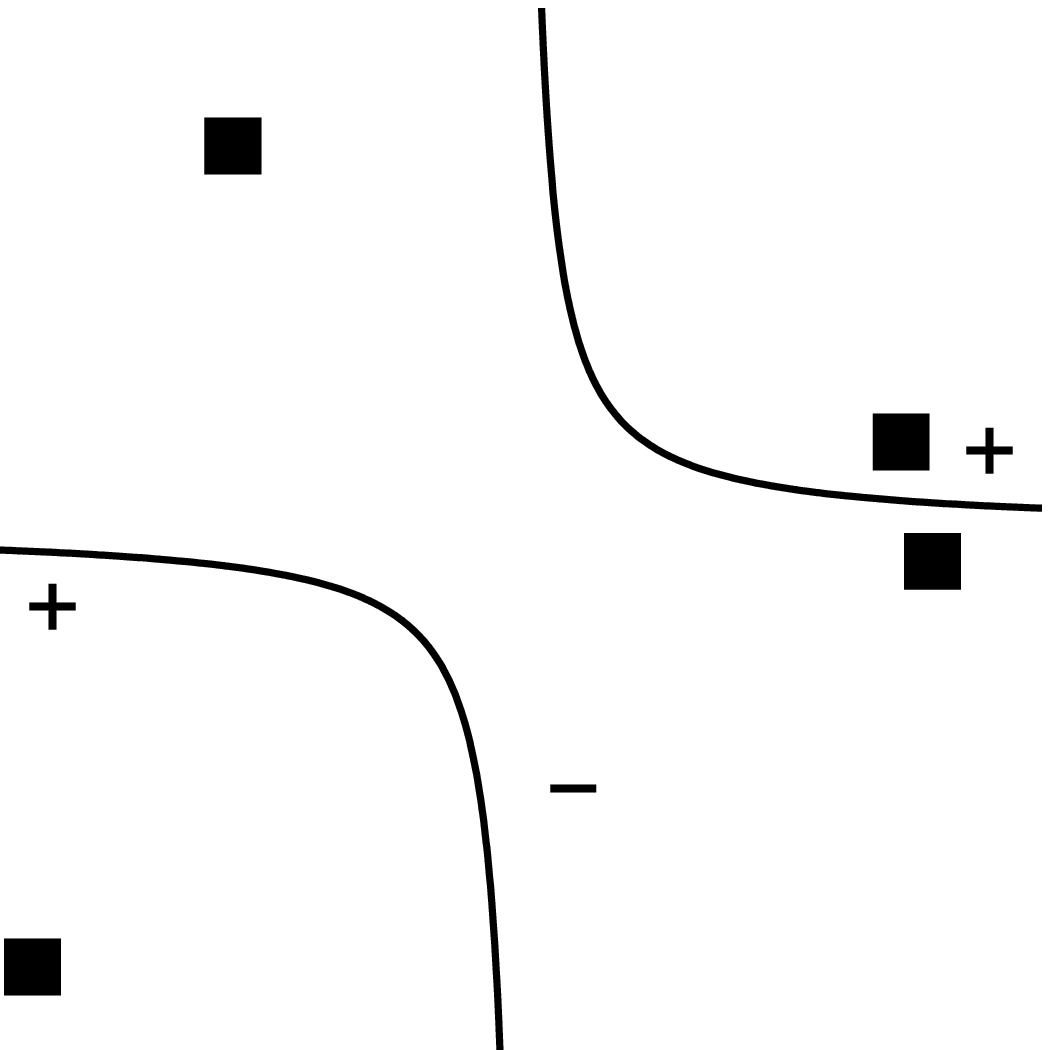}$~~~~~$\\
\\
\hline
\\
$~~~~~$\includegraphics[scale=.25]{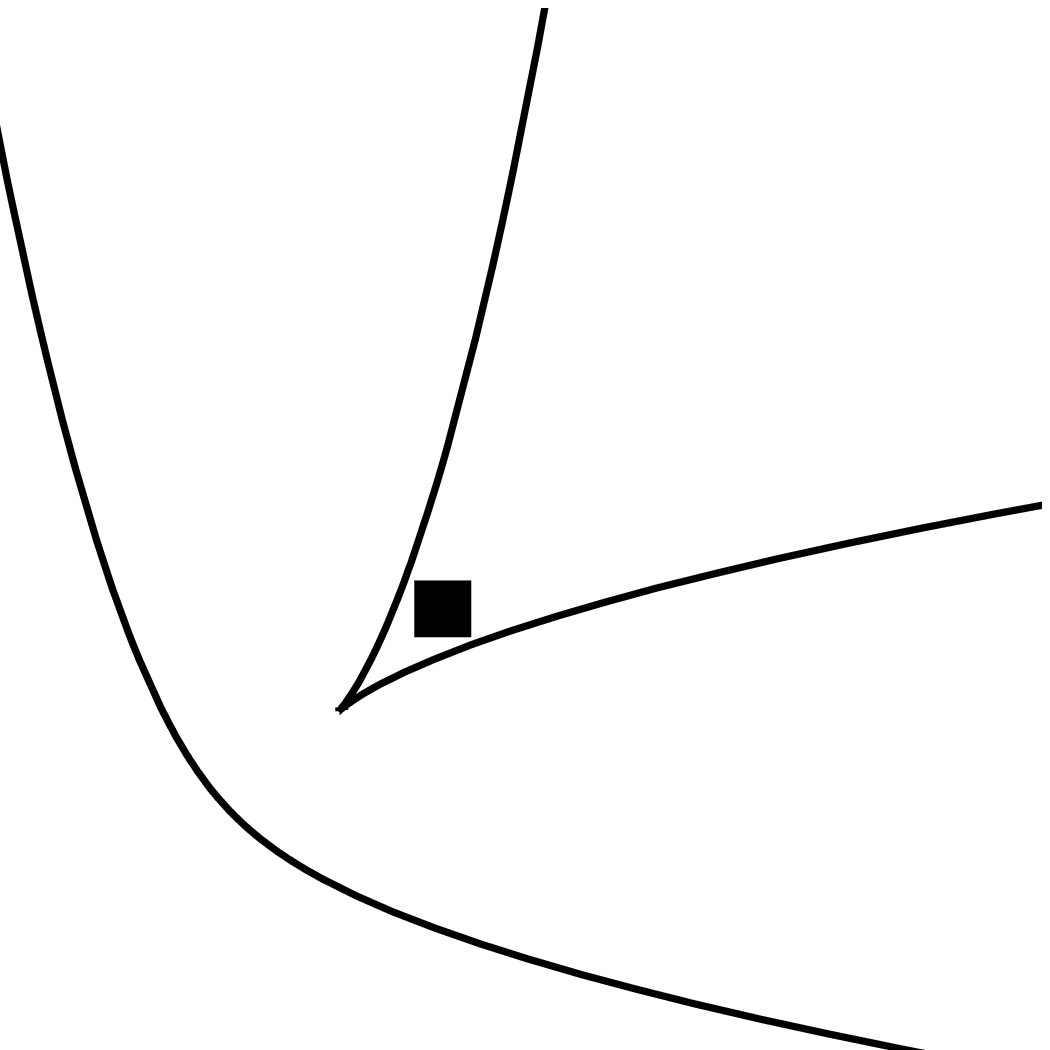}$~~~~~~~~~~$
\includegraphics[scale=.25]{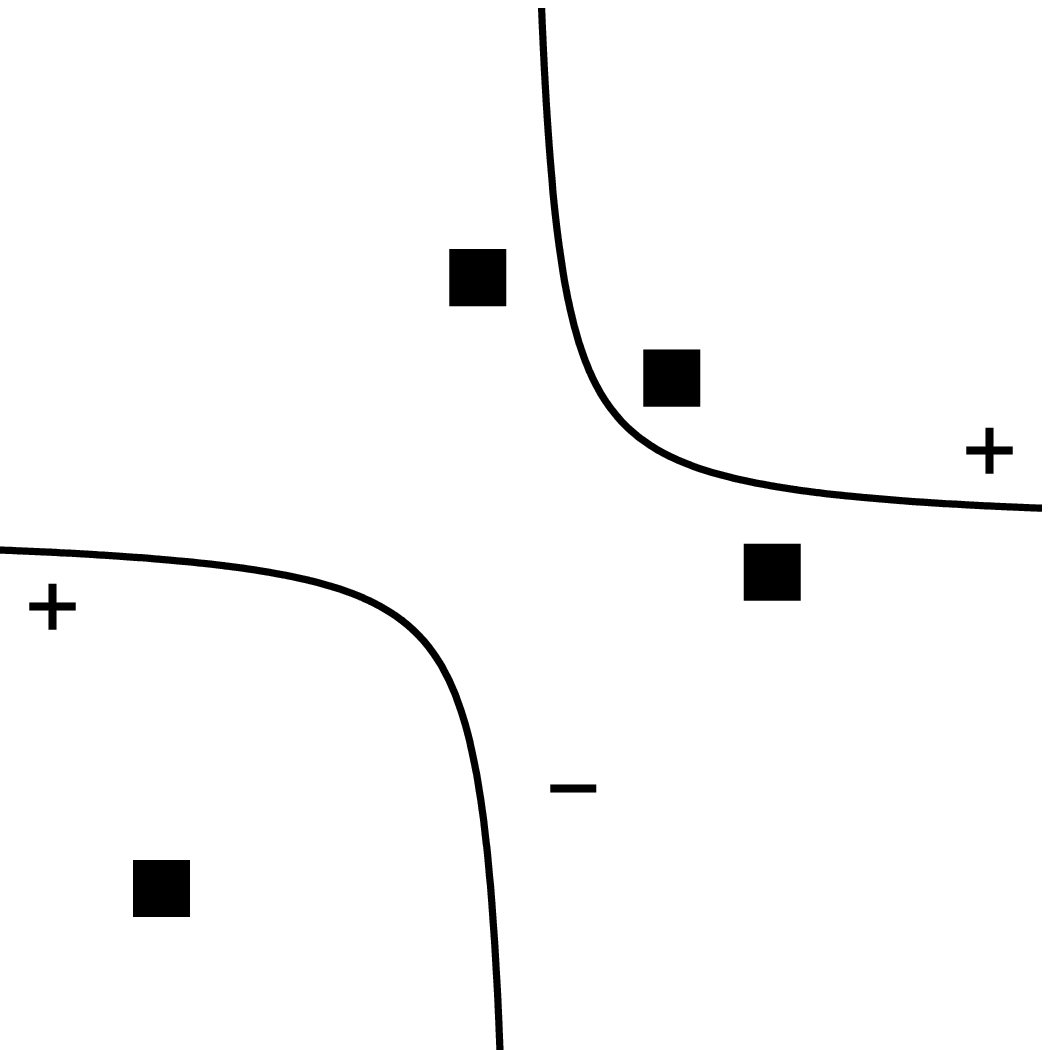}$~~~~~$\\
\\
\hline
\end{tabular}
\end{center}
\caption{Multiple images of a source inside a cross section of a
hyperbolic umbilic caustic in lensing.  In each panel, the figure on the left depicts the caustic curve with source position (solid box) in the light source plane, while the figure on the right shows the critical curve with corresponding image positions (solid boxes) in the lens plane.
The $+$ sign indicates that only minimum and maximum images lie
in the given region and the $-$ sign labels a region where only saddle images
occur.   Credits: Figure from \cite{Aazami-Petters1}.}
\label{fig:hyp}
\end{figure}

An example of the multiple imaging, critical curves, and caustic curves 
due to a hyperbolic umbilic ($D_4^{+}$) is shown in Figure~\ref{fig:hyp}.

\subsection{Universal Local Magnification Relations: Generic Caustics up to  Codimension 3}
\label{sec:univ-magrel-3}

\subsubsection{Preliminaries: Lensed Images and Magnification for Generic Mappings}

  Consider a smooth, $n$-parameter family $F_{{\bc},{\bs}}(\bx)$ of functions on an open subset of $\mathbb{R}^2$ that induces a smooth $(n-2)$-parameter family of mappings $\bbf_{\bc}(\bx)$ between planes ($n \geq 2)$.  The functions $F_{\bc,\bs}$ are used to construct a {\it Lagrangian submanifold} that is projected into the $n$-dimensional space $\{{\bc},{\bs}\} = \mathbb{R}^{n-2} \times \mathbb{R}^2$; the projection itself is called a {\it Lagrangian map}.  The critical values of this projection will then comprise the caustics of $\bbf_{\bc}$  (e.g.,  Golubitsky and Guillemin 1973 \cite{Gol-G}, Majthay 1985 \cite{Majthay}, Castrigiano and Hayes 1993 \cite{C-Hayes}, and \cite[pp. 276-86]{PLW}).  Arnold classified all stable simple Lagrangian map-germs of $n$-dimensional Lagrangian submanifolds by their generating family $F_{\bc,\bs}$ (\cite{Arnold73}, Arnold, Gusein-Zade, and Varchenko I 1985 \cite[p. 330-31]{AGV1}, and \cite[p. 282]{PLW}).

Given $\bbf_{\bc}(\bx) = \bs$, we call $\bx \in \RR^2$ a {\it lensed image} of the {\it source (or target) point} $\bs \in \RR^2$; note that our lensed images are never complex-valued.  Equivalently, lensed images are critical points of $F_{\bc,\bs}$, relative to a gradient in $\bx$.  Next, we define the {\it signed magnification} $\fkM(\bx_i;\bs)$ at a critical point $\bx_i$ of $F_{{\bc}, \bs}$ to be the reciprocal of the Gaussian curvature at the point $(\bx_i,F_{{\bc},\bs}(\bx_i))$ in the graph of $F_{{\bc},\bs}$:
\beqa
\label{Gauss}
\fkM({\bx_i; \bs})
= \frac{1}{{\rm Gauss}(\bx_i,F_{{\bc},\bs}(\bx_i))}\cdot \nonumber
\eeqa
The advantage of this definition is it makes clear that the signed magnification invariants are {\it geometric} invariants.  The signed magnification is expressible
in terms of $\bbf_{\bc}$ since
 ${\rm Gauss}(\bx_i,F_{{\bc},\bs}(\bx_i)) = {\rm det[Hess}\,F_{\bc,\bs}](\bx_i)$ and each $\bbf_{\bc}$ satisfies ${\rm det[Jac}\,\bbf_{\bc}] = {\rm det[Hess}\,F_{\bc,\bs}]$.  It follows that 
$$
\fkM (\bx_i; \bs) = \frac{1}{\det[\Jac \bbf_{\bc}](\bx_i)},
$$
 which is the more common definition of magnification.   If ${\rm det}[{\rm Jac}\,\bbf_{\bc}](\bx) = 0$, then $\bx$ is called a {\it critical point} of $\bbf_\bc$.  The collection of such points form curves called {\it critical curves}.  The target $\bbf_{\bc}(\bx)$ of a critical point $\bx$ is called a {\it caustic point}.  Though these typically form curves as well, they could also be isolated points.  Since there are caustic curves for each value of the parameters ${\bc}$, varying these parameters traces out a caustic surface, called a {\it big caustic,} in the larger space $\{\bs,\bc\} = \RR^{n}$.  An example of critical and caustic curves is shown in Figure~\ref{fig:hyp} for the hyperbolic umbilic ($D_4^{+}$), along with various source and image configurations.

\subsubsection{Universal Magnification Relations for Generic Caustics Up to Codimension 3}

The following theorem about
magnification relations for generic caustics up to codimension 3
was established by Aazami and AOP in 2009 \cite{Aazami-Petters1}:

\begin{theorem}{\rm \cite{Aazami-Petters1}}
\label{theorem:generic2}
For any of the  smooth generic families of functions $F_{{\bc},\bs}$ {\rm(}or the induced general mappings $\bbf_{\bc}${\rm)} giving rise to a caustic
of codimension up to $3$, and for $\bs$, any non-caustic point
in the $n$-image region, the following holds for the image magnifications
$\fkM_i \equiv  \fkM({\bx_i; \bs})$:
$$
\sum_{i=1}^n \fkM_i = 0,
$$
where 
$n =2$ for a fold, 
$n =3$ for a cusp,
$n =4$ for a swallowtail, elliptic umbilic, or hyperbolic umbilic.
\end{theorem}

The proofs of Theorems~\ref{theorem:quant} and \ref{theorem:generic2} in \cite{Aazami-Petters1},
including the subsequent shorter proof in \cite{Aazami-Petters2},
are algebraic in nature and the method will be highlighted in Section~\ref{sec:proof-ADEthm}.

After \cite{Aazami-Petters1} appeared, an alternative proof was given by MCW 2009 \cite{We09} that introduced new Lefschetz fixed point technology in gravitational lensing,
clarifying an earlier approach \cite{We07}.
The next section overviews the method in \cite{We09}.

\subsection{A Lefschetz Fixed Point Approach to Theorem~\ref{theorem:quant} and
Theorem~\ref{theorem:generic2}}

We first review some needed basics from Lefschetz fixed point theory and then outline
how it can be used to prove the aforementioned theorems.

\subsubsection{Holomorphic Lefschetz fixed point theory.} 
\label{sec:8.3.1}

If $f:M\rightarrow M$ is a smooth map on a compact manifold $M$, then its 
fixed points are $\fix(f)=\{x\in M:f(x)=x\}$, that is, the 
intersection of the graph $\{(x,f(x))\}\in M \times M$ with
the diagonal $\{(x,x)\}\in M \times M$. Fixed point theory, then, connects local 
properties of the fixed points, called fixed point indices, with global properties of $f$ and $M$. 
In the case of a real manifold $M$, this is called the {\it Lefschetz number} $L(f)$, which is a homotopy invariant 
because $f$ induces a map on the space of closed forms and hence on the cohomology classes of $M$. 
For complex $M$ and holomorphic $f$, the relationship between the analogous holomorphic Lefschetz number $L_\hol(f)$ 
and the local fixed point indices is called {\it holomorphic Lefschetz fixed point formula}. The Lefschetz fixed point formulas are well-defined provided that the intersections are transversal, and can be regarded as special cases of the Atiyah-Bott Theorem \cite{At67,At68}.

To illustrate this concept, we now discuss polynomial maps on the Riemann sphere $\hat{\C}=\C\PR^1=\C\cup\{\infty\}$ as an instructive example. Here, the holomorphic Lefschetz fixed point formula is also known as the Rational Fixed Point Theorem, which has important applications in complex dynamics (see, for example, the discussion by Milnor \cite{milnor}):
\begin{theorem}[Rational Fixed Point Theorem] Let $f: \hat{\C}\rightarrow\hat{\C}$ be a rational map which is not the identity. Then:
$$
1=\sum_{z\in \fix(f)}\frac{1}{1-\frac{\D f}{\D z}}.
$$
\end{theorem}
Here, the holomorphic Lefschetz number for complex projective space is $L_\hol(f)=1$. To see why this is true, recall that a rational map on the Riemann sphere can always be extended to a holomorphic map; then we can choose local holomorphic coordinates so that $z=0$ for some fixed point and write $f(z)=\frac{\D f}{\D z}(0) z + O(z^2)$ by holomorphicity. Hence,
\[
\frac{1}{(1-\frac{\D f}{\D z})(0)}=\frac{1}{2\pi \I}\oint_\sfC\frac{\D z}{z-f(z)},
\]
where $\sfC$ is a loop enclosing only the fixed point at the origin. We may take $f$ to be bounded so that 
\begin{equation}
\frac{1}{z-f(z)}-\frac{1}{z}=\frac{f(z)}{z(z-f(z))}\approx \frac{f(\infty)}{z^2}\rightarrow 0 \ \ \mbox{as} \ \ z\rightarrow \infty.
\label{integrand}
\end{equation}
Using these results, if we now extend the loop $\sfC$ to $\sfC_\infty$ of infinite radius enclosing all fixed points of $f$, then
\[
\sum_{z\in \fix(f)}\frac{1}{1-\frac{\D f}{\D z}}=
\frac{1}{2\pi \I}\oint_{\sfC_\infty}\frac{\D z}{z-f(z)}=
\frac{1}{2\pi \I}\oint_{\sfC_\infty}\frac{\D z}{z}=1 
\]
as required, since the integral of (\ref{integrand}) vanishes \cite{milnor}.

For a general complex manifold $M$ with dimension $d$, the holomorphic Lefschetz formula is (see Griffiths and Harris \cite{Gr78}, for instance)
\begin{equation}
L_\hol(f)=\sum_{z \in \fix(f)}\frac{1}{\det[I_d-D_f]},
\label{hololef}
\end{equation}
where $I_d$ is the $d$-dimensional identity matrix and $D_f$ is the matrix of first derivatives with respect to local holomorphic
coordinates. Again, the transversality condition can be expressed as the requirement that the fixed
point indices be well-defined, that is, $\det[I_d-D_f]\neq 0$.

\subsubsection{Application to Theorems~\ref{theorem:quant} and \ref{theorem:generic2}}
\label{sec:8.3.2}

\begin{table}
\centering
\begin{tabular}{llll}
\hline\noalign{\smallskip}
Singularity & $\fo,\ft$ & $\deg(\fo), \deg(\ft)$ & $n$  \\
\noalign{\smallskip}\hline\noalign{\smallskip}
Fold & $z_1, z_2^2$ & $1, 2$ & $2$ \\
Cusp & $z_1, z_1z_2+z_2^3$ & $1, 3$ & $3$ \\
Swallowtail & $z_1z_2+cz_1^2+z_1^4, z_2$ & $4, 1$ & $4$ \\
Elliptic umbilic & $3z_2^2-3z_1^2-2cz_1, 6z_1z_2-2cz_2$ & $2, 2$ & $4$ \\
Hyperbolic umbilic  & $-3z_1^2-cz_2, -3z_2^2-cz_1$ & $2, 2$ & $4$ \\
\noalign{\smallskip}\hline\noalign{\smallskip}
Elliptic umbilic (lensing map) & $z_1^2-z_2^2, -2z_1z_2+4cz_2$ & $2, 2$ & $4$ \\
Hyperbolic umbilic (lensing map) & $z_1^2+2cz_2, z_2^2+2cz_1$ & $2, 2$ & $4$ \\
\noalign{\smallskip}\hline
\end{tabular}
\caption{Components of the complex maps $\f(z_1,z_2)$ for generic singularities up to codimension three, their degrees and the numbers $n$ of real solutions in the maximal caustic domains at finite positions. Here $c$ denotes a control parameter. The lower section lists the corresponding properties for 
the quantitative elliptic and hyperbolic umbilics for a lensing map.}
\label{tab:complex}
\end{table}

We now outline the Lefschetz fixed point proof of Theorem~\ref{theorem:generic2}.
In the present approach
(compare Atiyah and Bott \cite{At68}), all real solutions $\mathbf{f}_c(x_1,x_2)=(s_1,s_2)$ are treated as the real fixed points of a suitable complex map. So, first, we need to find a complexification that allows the application of the holomorphic Lefschetz fixed point formula. The standard complexification $(x_1,x_2)\mapsto x_1+\I x_2$ does not yield holomorphic maps, but this problem can be circumvented, at the expense of the dimension, by treating $(x_1,x_2)\equiv(z_1,z_2)$ as independent complex variables on $\C^2$. The corresponding complex generic maps $\f=(\fo,\ft):\C^2\rightarrow \C^2$ are shown in Table \ref{tab:complex}, and the maximum number of solutions of $\fo(z_1,z_2)=s_1,\ft(z_1,z_2)=s_2$, possibly complex, is $\deg(\fo)\deg(\ft)$ by B\'ezout's Theorem. Now, since Table \ref{tab:complex} shows that this is always equal to the maximum number $n$ of real and finite solutions, we see that our complex formalism gives the usual real solutions in the maximal caustic domain, as required. Next, one can define the map $f=(f_1,f_2)=(z_1-\fo+y_1,z_2-\ft+y_2)$ on $\C^2$ such that its fixed points are in fact those solutions. Hence, by construction, $f$ is holomorphic and has no fixed points at infinity. Notice also that, at these fixed points,

\begin{equation}
\frac{1}{\det[I_2-D_f]}=\frac{1}{\det [\mathrm{Jac} \ \mathbf{f}_c]}=\mathfrak{M}.
\label{mag}
\end{equation}
Although this looks rather suggestive, we cannot apply the holomorphic Lefschetz fixed point formula directly to $f$, since $\C^2$ is not compact. However, one can rewrite $f$ in homogeneous coordinates $(Z_0,Z_1,Z_2)$, where $z_1=Z_1/Z_0$ and $z_2=Z_2/Z_0$ for $Z_0\neq0$, and consider the a map $F=(F_0:F_1:F_2)$ on $\C\PR^2$, which is of course compact, where
\begin{eqnarray*}
F_0&=&Z_0^m, \\
F_1&=&Z_1 Z_0^{m-1}-Z_0^{m-\deg(\fo)}\fo(Z_0,Z_1,Z_2)+y_1 Z_0^m, \\
F_2&=&Z_2 Z_0^{m-1}-Z_0^{m-\deg(\ft)}\ft(Z_0,Z_1,Z_2)+y_2 Z_0^m,
\end{eqnarray*}
and $m=\max(\deg(\fo),\deg(\ft))\geq 2$. Thinking of $\C\PR^2=\C^2 \cup \C\PR^1$, with $\C^2: Z_0=1, \ \C\PR^1: Z_0=0$, we recover $F|_{\C^2}=f$. Also, $F$ is holomorphic since $f$ has no fixed points at infinity. Finally, the holomorphic Lefschetz fixed point formula (\ref{hololef}) applies to $F$ and to $F|_{\C\PR^1}$, where $L_\hol=1$ as shown in Section~\ref{sec:8.3.1}. Hence, using (\ref{mag}), we obtain $1=\sum_{i=1}^n \mathfrak{M}_i+1$. This shows that the universal magnification invariant is zero for the generic cases in Theorem~\ref{theorem:generic2}, which can now be interpreted as a difference of two Lefschetz numbers. 

For the quantitative case a similar argument yields a proof of Theorem~\ref{theorem:quant};
the lower section of Table~\ref{tab:complex} shows that the
properties of the quantitative elliptic and hyperbolic umbilics for
lensing maps allow a construction as discussed in Section~\ref{sec:8.3.2}, in exactly the same way as for the corresponding generic maps.

\smallskip

\noindent
{\bf Remark:}
The above fixed point approach to lensing should not be confused
with the one studied by AOP and Wicklin 1998 \cite{petters98}, where fixed points
of the lensing map were explored to determine source positions that have
a lensed image coinciding with the unlensed angular position of the source.
 It is also interesting to note that the technical transversality
condition mentioned in Section~\ref{sec:8.3.1} simply becomes
$|\mu|<\infty$ in the lensing context, that is, the usual condition for
regular images.

\section{Universal Local Magnification Relations: Generic Caustics Beyond Codimension 3}
\label{sec:univ-magrel-0}

\subsection{The Infinite Family of A, D, E Caustics}
\label{sec:8.2.1}

In Arnold's classification of stable simple Lagrangian map-germs of $n$-dimensional Lagrangian submanifolds by their generating family $F_{\bc,\bs}$ \cite{Arnold73}
(also, see Arnold, Gusein-Zade, and Varchenko \cite[p. 330-31]{AGV1}),
he found a deep connection between his classification and the Coxeter-Dynkin diagrams of the simple Lie algebras of types $A_n~(n \geq 2), D_n~(n \geq 4), E_6, E_7, E_8$.  This classification is shown in Table~\ref{table:ADE} \cite{Aazami-Petters3} and is known as the {\it A, D, E classification} of caustic singularities.

\begin{table}
\centering
\begin{tabular}{ll}
\hline\noalign{\smallskip}

Class & $F_{\bc,\bs}(x_1,x_2)$ \\
\mbox{} & ${\bf f}_{\bc}(x_1,x_2)$ \\
 
\noalign{\smallskip}\hline\noalign{\smallskip}

{\it \large{A}}$_{n}$ & ${\pm x_1^{n+1} \pm x_2^2} + {c_{n-1}x_1^{n-1} +
\cdots + c_3x_1^3 + s_2x_1^2 - s_1x_1 \pm s_2x_2}$ \\

$(n \geq 2)$ & $\left(\pm(n+1)x_1^{n} + (n-1)c_{n-1}x_1^{n-2} + \cdots +
3c_3x_1^2 \mp 4x_2x_1~,~\mp2x_2\right)$ \\

\noalign{\smallskip}\hline\noalign{\smallskip}

{\it \large{D}}$_n$ & ${x_1^2x_2 \pm x_2^{n-1}} + {c_{n-2}x_2^{n-2} +
\cdots + c_2x_2^2 - s_2x_2 - s_1x_1}$ \\

$(n \geq 4)$ & $\left(2x_1x_2~,~x_1^2\pm(n-1)x_2^{n-2} +
(n-2)c_{n-2}x_2^{n-3} + \cdots + 2c_2x_2\right)$ \\

\noalign{\smallskip}\hline\noalign{\smallskip}

{\it \large{E}}$_6$ & ${x_1^3 \pm x_2^4} + {c_3x_1x_2^2 + c_2x_2^2 +
c_1x_1x_2 - s_2x_2 - s_1x_1}$ \\

\mbox{} & $\Big(3x_1^2 + c_3x_2^2 +c_1x_2~,~\pm4x_2^3 + 2c_3x_1x_2 +
2c_2x_2 + c_1x_1\Big)$ \\

\noalign{\smallskip}\hline\noalign{\smallskip}

{\it \large{E}}$_7$ & ${x_1^3 + x_1x_2^3} + {c_4x_2^4 + c_3x_2^3 +
c_2x_2^2 + c_1x_1x_2 - s_2x_2 - s_1x_1}$ \\

\mbox{} & $\Big(3x_1^2 + x_2^3 + c_1x_2~,~3x_1x_2^2 + 4c_4x_2^3 +
3c_3x_2^2 + 2c_2x_2 + c_1x_1\Big)$ \\

\noalign{\smallskip}\hline\noalign{\smallskip}

{\it \large{E}}$_8$ & ${x_1^3 + x_2^5} + {c_5x_1x_2^3 + c_4x_1x_2^2 +
c_3x_2^3 + c_2x_2^2 + c_1x_1x_2 - s_2x_2 - s_1x_1}$ \\

\mbox{} & $\Big(3x_1^2 + c_5x_2^3 + c_4x_2^2 + c_1x_2~,~5x_2^4 +
3c_5x_1x_2^2 + 2c_4x_1x_2 + 3c_3x_2^2 + 2c_2x_2 +c_1x_1\Big)$ \\

\noalign{\smallskip}\hline
\end{tabular}
\caption{The left column indicates the $A, D, E$ type of the Coxeter-Dynkin diagram or generic caustic.
The right  column lists the associated universal local forms of the smooth $(n-1)$-parameter family of general functions $F_{{\bo c},\bs}$, along with their $(n-3)$-parameter family of induced general maps $\bbf_{\bo c}$ between planes---see 
the two-component expressions
$(...\ , \ ...)$.
 The given classification is due to Arnold 1973 \cite{Arnold73}. 
Credits: Table from \cite{Aazami-Petters3}.
}
\label{table:ADE}
\end{table}

The generic caustic singularities up to codimension $5$ are given as follows
in the Arnold $A, D, E$ notation,
where the numbers nD in parentheses indicate the codimension:
\begin{enumerate}
 \item (1D)\  $A_2$ is a fold.
 \item  (2D)\ $A_3$ is a cusp.
\item (3D)\ $A_4$ is a swallowtail, \ $D^-_4$ an elliptic umbilic, and 
           $D^+_4$ a hyperbolic umbilic.
\item {\rm (4D)}\ $A_5$ is a butterfly, \ $D_5$ a parabolic umbilic.
\item {\rm (5D)}\ $A_6$ is a wigwam, \ $D_6^-$ a 2nd elliptic umbilic, \ $D_6^+$ a 2nd hyperbolic umbilic,
               \ $E_6$  symbolic umbilic.
\end{enumerate}
\smallskip

\noindent
{\bf Remark:} Up to codimension $5$,
all Lagrangian maps can be
approximated by stable Lagrangian map-germs \cite{Arnold73}.
However, for codimension $6$ or higher,  this is no longer the case;
unstable mappings form an open dense set \cite{Arnold73,AGV1}.

\subsection{Universal Magnification Relations for the Family of A, D, E Caustics}
\label{sec:univ-magrel}

In 2009, Aazami and AOP  \cite{Aazami-Petters3}
proved  a  univeral local magnification relation theorem for 
generic general mappings between planes exhibiting {\it any} caustic singularity appearing in Arnold's  $A, D, E$ family:

\begin{theorem}{\rm \cite{Aazami-Petters3}}
\label{theorem:generic}
For any of the generic smooth $(n-1)$-parameter family of general functions $F_{{\bc},\bs}$ 
{\rm(}or induced general mappings $\bbf_{\bc}${\rm)} in the A, D, E classification, and for any non-caustic point $\bs$ in the indicated region, the following results hold for the magnification $\fkM_i \equiv  \fkM({\bx_i; \bs})$:
\begin{enumerate}
\item $A_n~(n \geq 2)$ obeys the magnification relation in the $n$-image region: $\sum_{i = 1}^{n} \fkM_i = 0,$
\item $D_n~(n \geq 4)$ obeys the magnification relation in the $n$-image region: $\sum_{i = 1}^{n} \fkM_i = 0,$
\item $E_6$ obeys the magnification relation in the six-image region: $\sum_{i = 1}^{6} \fkM_i = 0,$
\item $E_7$ obeys the magnification relation in the seven-image region: $\sum_{i = 1}^{7} \fkM_i = 0,$
\item $E_8$ obeys the magnification relation in the eight-image region: $\sum_{i = 1}^{8} \fkM_i = 0.$
\end{enumerate}
\end{theorem}
\smallskip

\noindent
{\bf Remark:} Theorem~\ref{theorem:generic} does not 
follow directly from the Euler-Jacobi formula, the multi-dimensional residue integral
approach \cite{Dalal-Rabin}, or the Lefschetz fixed point theory
method \cite{We09},  because some of the singularities have fixed points at infinity.

\subsection{On the Proof of Theorem~\ref{theorem:generic}}
\label{sec:proof-ADEthm}

The proof of Theorem~\ref{theorem:generic} given in \cite{Aazami-Petters3} 
employed the Euler trace formula. This formula
was shown by Aazami and AOP \cite{Aazami-Petters2}
in 2009 to be a corollary of a more general result they established
about polynomials:

\begin{theorem}{\rm \cite{Aazami-Petters2}}
\label{prop:recursive}
Let $\varphi(x) = a_n x^n + \cdots + a_1 x + a_0 \in \mathbb{C}[x]$ be any polynomial with distinct roots $x_i$, and let $h(x)\in R$ be any rational function, where $R$ is the subring of rational functions that are defined at the roots of $\varphi(x)$.  Let 
$$
h_*(x) = c_{n-1} x^{n-1} + \cdots + c_1 x + c_0
$$
be the unique polynomial representative of the coset $\oli{h(x)} \in \rideal$ and let
$$
r(x) = b_{n-1} x^{n-1} + \cdots + b_1 x + b_0
$$
be the unique polynomial representative of
the coset $\oli{\varphi'(x)h(x)}\in \rideal$.
Then the coefficients of $r(x)$ are
given in terms of the coefficients of $h_*(x)$ and $\varphi(x)$ 
by the following recursive relation:
\beqa
\label{eq:gen-recursive}
b_{n-i} = c_{n-1} b_{n-i,n-1} + \cdots + c_1 b_{n-i,1} + c_0 b_{n-i,0} \,
\hspace{0.75in} i = 1, \dots, n\ , \nonumber
\eeqa
with
\beqa
\label{eq:relations}
\left\{
\begin{array}{ll}
b_{n-i, 0} = (n- (i-1))\, a_{n- (i-1)}\ , & \qquad i = 1, \dots, n\ ,\\
                                       & \\
\displaystyle b_{n-i,k} = -\frac{a_{n-i}}{a_{n}}\, b_{n-1,k-1} + b_{n-(i+1),k-1}\ , & 
                 \qquad  i = 1, \dots, n\ , \qquad  k = 1, \dots, n-1\ ,\nonumber
\end{array}
\right.
\eeqa
where $b_{-1,k-1} \equiv 0$.
\end{theorem}
\smallskip

\begin{corollary}[Euler Trace Formula]
Assume the hypotheses and notation of Theorem~\ref{prop:recursive}.  For any rational function $h(x) \in R$,  the following holds:
\beqa
\label{euler}
\sum_{i=1}^{n} h(x_i)  = \frac{b_{n-1}}{a_n}\cdot\nonumber
\eeqa
\end{corollary}
\noindent
See \cite{Dalal-Rabin} for a residue calculus approach to the Euler trace formula.

One can now show that the total signed magnification satisfies:
\beqa
\label{eq:magsum}
\sum_{i} \fkM_i = \frac{b_{n-1}}{a_{n}}\cdot
\eeqa
For all of the caustic singularities appearing in the infinite family of $A_n~(n \geq 2), D_n~(n \geq 4), E_6, E_7, E_8$ singularities, the coefficient $b_{n-1}$ was shown to be zero, and Theorem~\ref{theorem:generic} was thereby proved.  We will illustrate the method of proof here in the case of the hyperbolic umbilic.  See \cite{Aazami-Petters2,Aazami-Petters3} for a detailed treatment.

The induced map $\bbf_c$ corresponding to the hyperbolic umbilic is
given by:
\beqa
\label{hyp}
{\bbf_c}(x_1,x_2) = (-3 x_1^2 - c x_2\,,\,-3 x_2^2 - c x_1).\nonumber
\eeqa
Let $\bs = (s_1,s_2)$ be a target point lying in the four-image region.
 The four lensed images of $\bs$ are obtained by solving for the equation
\beqa
\label{hyp:pol}
(-3 x_1^2 - c x_2\,,\,-3 x_2^2 - c x_1) = (s_1,s_2).
\eeqa
To use the Euler trace formula in the form (\ref{eq:magsum}), we begin
by eliminating $x_2$ to obtain a polynomial in the variable $x_1$:
$$
\varphi(x_1) \equiv -3 s_1^2 - c^2 s_2 - c^3 x_1 - 18 s_1 x_1^2 - 27 x_1^4.
$$
The magnification of a lensed image of $\bs$ under $\bbf_c$ is
$\fkM(x_1,x_2) = 1/(-c^2 + 36 x_1 x_2)$.  To convert this into a
rational function in the single variable $x_1$, we substitute for $x_2$
via (\ref{hyp:pol}) to obtain:
$$
 \fkM(x_1,x_2(x_1)) = \frac{c}{-c^3 - 36 s_1 x_1 - 108 x_1^3} \equiv
\fkM(x_1).
$$
A direct calculation now yields:
$$
\varphi'(x_1)\fkM(x_1) = c.
$$
Thus the unique polynomial representative in the coset
$\oli{\varphi'(x_1)\fkM(x_1)}$ is the polynomial $r(x_1) \equiv c$ (in
the notation of Theorem~\ref{prop:recursive}, $\fkM(x_1) \equiv
h(x_1)$).  Since $b_{n-1} = b_3 = 0$, (\ref{eq:magsum}) tells us
immediately that
$$
\fkM_1 + \fkM_2 + \fkM_3 + \fkM_4 = 0.
$$

\begin{acknowledgements}
AOP and MCW would like to thank Amir Aazami and Alberto Teguia for stimulating discussions,
and Stanley Absher for editorial assistance. 
\end{acknowledgements}



\end{document}